\documentclass[%
 aip,
 amsmath,amssymb,
reprint,%
]{revtex4-1}

\usepackage[pdftex]{color,graphicx}
\usepackage{dcolumn}% Align table columns on decimal point
\usepackage{bm}% bold math
\usepackage[utf8]{inputenc}
\usepackage[T1]{fontenc}
\usepackage{mathptmx}
\usepackage{siunitx}

\usepackage[english]{babel}
\usepackage{hyperref}
\newcommand{\figref}[1]{Fig.~\ref{#1}}

\makeatletter
\DeclareRobustCommand{\change}{%
  \@bsphack
  \leavevmode
  \@esphack
}
\DeclareRobustCommand{\stopchange}{%
  \@bsphack
  \@esphack
}
\makeatother

\graphicspath{{./img/}}
\DeclareGraphicsExtensions{.pdf,.png,.jpg}
\DeclareUnicodeCharacter{00A0}{~}

\newcommand{\PDCKTH}{Swedish e-Science Research Center, PDC Center for High Performance Computing, KTH Royal Institute of Technology,
100 44 Stockholm, Sweden}
\newcommand{\AppPhysKTH}{Department of Applied Physics, Swedish e-Science Research Center, Science for Life Laboratory, KTH Royal Institute of Technology, Box 1031, 171 21 Solna, Sweden}
\newcommand{\DbbSU}{Department of Biochemistry and Biophysics, Science for Life Laboratory, Stockholm University, Box 1031, 171 21 Solna, Sweden}

\begin{document}
%\preprint{AIP/123-QED}


\title[]{Heterogeneous Parallelization and Acceleration of Molecular Dynamics Simulations in GROMACS}

\renewcommand*{\thefootnote}{\fnsymbol{footnote}}

\author{Szil\'ard P\'all}
  \affiliation{\PDCKTH}
\author{Artem Zhmurov}
  \affiliation{\PDCKTH}
\author{Paul Bauer}
    \affiliation{\AppPhysKTH}
\author{Mark Abraham}
    \affiliation{\AppPhysKTH}
\author{Magnus Lundborg}
    \affiliation{ERCO Pharma AB, Stockholm, Sweden}
\author{Alan Gray}
    \affiliation{NVIDIA Corporation, UK}
\author{Berk Hess}
\thanks{To whom correspondence should be adressed; hess@kth.se, erik.lindahl@scilifelab.se}
    \affiliation{\AppPhysKTH}
\author{Erik Lindahl}
\thanks{To whom correspondence should be adressed; hess@kth.se, erik.lindahl@scilifelab.se}
   \affiliation{\AppPhysKTH}
    \affiliation{\DbbSU}
%%%%%%%%%%%%%%%%%%%%%%%%%%%%%%%%%%%%%%%%%%%%%%%%%%%%%%%%%%%%%%%%%%%%%%%%%%%%%%%%%%%%%%%%%%

%%%%%%%%%%%%%%%%%%%%%%%%%%%%%%%%%%%%%%%%%%%%%%%%%%%%%%%%%%%%%%%%%%%%%%%%%%%%%%%%%%%%%%%%%%
%% Top-level / general TODO list
%
% Minor:
% - use \SI{}{}
% - go through all definitions () as last thing
% - Make sure that "--" is used for ranges
% - Footnotes are not in-page, fix that
%
%%%%%%%%%%%%%%%%%%%%%%%%%%%%%%%%%%%%%%%%%%%%%%%%%%%%%%%%%%%%%%%%%%%%%%%%%%%%%%%%%%%%%%%%%%

\date{\today}

\begin{abstract} \label{sec:abstract}
% Abstract is DONE.
The introduction of accelerator devices such as graphics processing units (GPUs) has had profound impact on molecular dynamics simulations and has enabled order-of-magnitude performance advances using
commodity hardware. To fully reap these benefits, it has been necessary to reformulate
some of the most fundamental algorithms, including the Verlet list, pair searching and cut-offs. 
Here, we present the heterogeneous parallelization and acceleration design of molecular dynamics
implemented in the GROMACS codebase over the last decade. The setup involves
a general cluster-based approach to pair lists and non-bonded pair interactions that utilizes both
GPUs and CPU SIMD acceleration efficiently, including the ability to load-balance tasks between CPUs and GPUs. The algorithm work efficiency is tuned for each type of hardware, and to use accelerators more
efficiently we introduce dual pair lists with rolling pruning updates. Combined with new direct GPU-GPU
communication as well as GPU integration, this enables excellent performance from single GPU
simulations through strong scaling across multiple GPUs and efficient multi-node parallelization. 
\end{abstract}

\maketitle

\section{Introduction} \label{sec:intro}
Molecular dynamics (MD) simulation has had tremendous success in a number
of application areas the last two decades, in part due to hardware improvements
that have enabled studies of systems and timescales
that were previously not feasible. These advances have also made it possible
to introduce better algorithms and longer simulations have enabled
more accurate calibration of force fields against experimental data, all of which
have contributed to increasing trust in computational studies. However, 
the high computational cost of evaluating forces between all particles 
combined with integrating over short time steps (${\sim} \SI{2}{\femto \second}$) has led to 
fundamental challenges for the field as the speed of individual processor cores
is no longer increasing. Without algorithms that can better exploit new parallel hardware,
the timescales accessible in simulations will hit a brick wall. 
Unlike some other fields, improving resolution by increasing the model detail,
e.g.\ with quantum effects or increasing the size of the system, cannot replace
reaching longer timescales. In most cases molecular dynamics targeting a specific
application depends critically on achieving faster simulations by reducing the time
each MD step takes. 

One successful recent approach has been the introduction of enhanced
sampling based on ensembles of simulations. When combined with parallelization of individual runs,
this makes it possible to use the largest high performance computing (HPC) resources
in the world to study even small systems. However, even for HPC systems a high rate of producing trajectories is imperative to sample dynamics covering adequate timescales, which means cost-efficiency and throughput are of paramount importance\cite{Kutzner2019}.

The design choices in GROMACS are guided by a bottom-up approach to parallelization and optimization,
partly due to the code's roots of high performance on cost-efficient hardware.  This is
not without challenges; good arguments can be made for focusing either top-down on
scaling or just sticking to single-GPU simulations. However, by employing state-of-the-art
algorithms and efficient parallel implementations the code is able to target hardware
and efficiently parallelize from the lowest level of SIMD (single instruction, multiple data)
vector units to multiple cores and caches, accelerators, and distributed-memory HPC resources.

We believe this approach makes great use of limited compute resources to improve research
productivity\cite{Loeffler2012,Schaffner2018}, and it is increasingly enabling higher absolute
performance on any given resource. Exploiting low-level parallelism can be tedious and
has often been avoided in favor of using more hardware to achieve the desired time-to-solution.
However, the evolution of hardware is making this trade-off increasingly
difficult. The end of microprocessor frequency scaling and the consequent increase in
hardware parallelism means that targeting all levels of parallelism is a necessity rather than
option. The MD community has been at the forefront of investing in this direction\cite{Yoshii2014,Brown2016,McDoniel2017,Acun2018}, and our early work 
on scalable algorithms\cite{Hess2008}, fine-grained
parallelism\cite{Pall2013} and low-level portable parallelization abstractions\cite{Abraham2015}
have been previous steps on this path.

Accelerators such as GPUs are expected to provide the majority of raw FLOPS in upcoming
Exascale machines. However, the impact of GPUs can also be seen in low- to mid-range capacity computing, especially in fields like MD that have been able to utilize the high instruction throughput as well
as single precision capabilities; this has had particularly high impact in making consumer GPU
hardware available for scientific computing.

While algorithms with large amounts of fine-grained parallelism are well-suited to GPUs, 
tasks with little parallelism or irregular data-access are better suited
to CPU architectures. Accelerators have become increasingly flexible, but still require 
host systems equipped with a CPU. While there has been some convergence of architectures, 
the difference between latency- and throughput-optimized functional units is fundamental,
and utilizing each of them for the tasks at which they are best suited requires
\emph{heterogeneous parallelization}. This typically employs the CPU also
for scheduling work, transferring data and launching computation on the accelerator,
as well as inter- and intra-node communication. Accelerator tasks are launched
asynchronously using APIs such as CUDA, OpenCL or SYCL to allow concurrent CPU--GPU execution. 
The heterogeneous parallelization model adds complexity which comes at a cost, both in terms of hardware
(latency, complex topology) and programmability, but it provides flexibility (every single
algorithm does not need to be implemented on the accelerator) and opportunities for
higher performance. Heterogeneous systems are evolving fast with very
tightly coupled compute units, but the heterogeneity in HPC will remain and is likely
best addressed explicitly.

Our first GPU support was introduced in GROMACS 4.5\cite{Pronk2013}
and relied on a homogeneous acceleration by implementing the entire MD calculation
on the GPU. The same approach has been used by several codes
(e.g.\ ACEMD\cite{Harvey2009a}, AMBER\cite{Gotz2012, Salomon-Ferrer2013},
HOOMD-blue\cite{Anderson2008}, FENZI\cite{Ganesan2011a},
DESMOND-GPU\cite{Bergdorf2016})
and has the benefit that it keeps the GPU busy,
avoiding communication as long as scaling is not a concern. However, this first 
approach also had shortcomings: only algorithms ported to the GPU can be used
in simulations, which limits applicability in large community codes. Implementing the full set of MD
algorithms on all accelerator frameworks is not practical from porting and maintenance concerns.
In addition, our experience showed that outperforming the highly optimized CPU code in GROMACS by only
relying on GPUs was difficult, especially in parallel runs where the CPU-accelerated code excels.
To make use of GPUs without giving up feature support, while providing speedup to as many simulation
use-cases as possible, utilizing both CPU and GPU in heterogeneous parallelization was necessary.

Heterogeneous offload is employed by several MD codes
(NAMD \cite{Stone2007}, LAMMPS \cite{Brown2012}, CHARMM \cite{Hynninen2013}, or GENESIS \cite{Kobayashi2017}).
However, here too the GROMACS CPU performance provided a challenge:
since the tuned CPU SIMD kernels are already capable of achieving iteration rates around 
\SI{1}{\milli \second} per step without GPU acceleration, the {\em relative} speedup of adding an accelerator was
less impressive at the time.

To address this, we started from scratch by recasting algorithms into a 
future-proof form to exploit both GPUs and CPUs (including multiple devices)
to provide very high absolute performance while supporting virtually all features no matter what
hardware is available. The pair-interaction calculation was redesigned with a cluster pair algorithm\cite{Pall2013} to fit modern architectures, which replaces the traditional Verlet
list based on particles. Clusters are optimized to fit the hardware, and the classical cut-off setup
has evolved into accuracy-based approaches for simulation settings to allow multi-level load
balancing and on-the-fly tuning based on system properties. Together with CPU SIMD parallelization and
multi-threading\cite{Pall2014}, this has allowed efficient offloading of short-range non-bonded
calculations to GPUs and brought major gains in performance\cite{Kutzner2013}.

New algorithms and the heterogeneous acceleration framework have made it possible to track
track the shift towards dense heterogeneous machines and balancing CPU/GPU utilization by
offloading more work to powerful accelerators.\cite{Kutzner2019}
The most recent release has almost come full circle to allow offloading full MD steps, but this
version also supports most features of the MD engine by utilizing both CPUs and GPUs, it targets multiple accelerator architectures, and provides scaling both across multiple accelerator devices and multiple nodes.
This bottom-up heterogeneous acceleration approach provides flexibility, portability
and performance for a wide range of target architectures, ranging from laptops to supercomputers and
from CPU-only machines to dense multi-GPU clusters. Below we present the algorithms and implementations 
that have enabled it.
%%% Szilard: added per request from R#2
\change
These are relatively complex concepts, so to aid the reader we will first discuss the general ideas in sections II-IV,
after which we dedicate section V to details of core algorithms, returning to performance benchmarks and discussions in the final two sections.
\stopchange

%%%%%%%%%%%%%%%%%%%%%%%%%%%%%%%%%%%%%%%%%%%%%%%%%%%%%%%%%%%%%%%%%%%%%%%%%%%%%%%%%%%%%%%%%%
\section{Computational challenges in MD simulations}\label{sec:challenges-in-MD}
The core of classical MD is the time-evolution of particle
systems by numerically integrating Newton’s equations of motion. This requires calculating forces
for every time step, which is the main computational cost. While this
can be parallelized, the integration step is inherently iterative.

\begin{figure}
\center
\includegraphics[width=0.95\columnwidth]{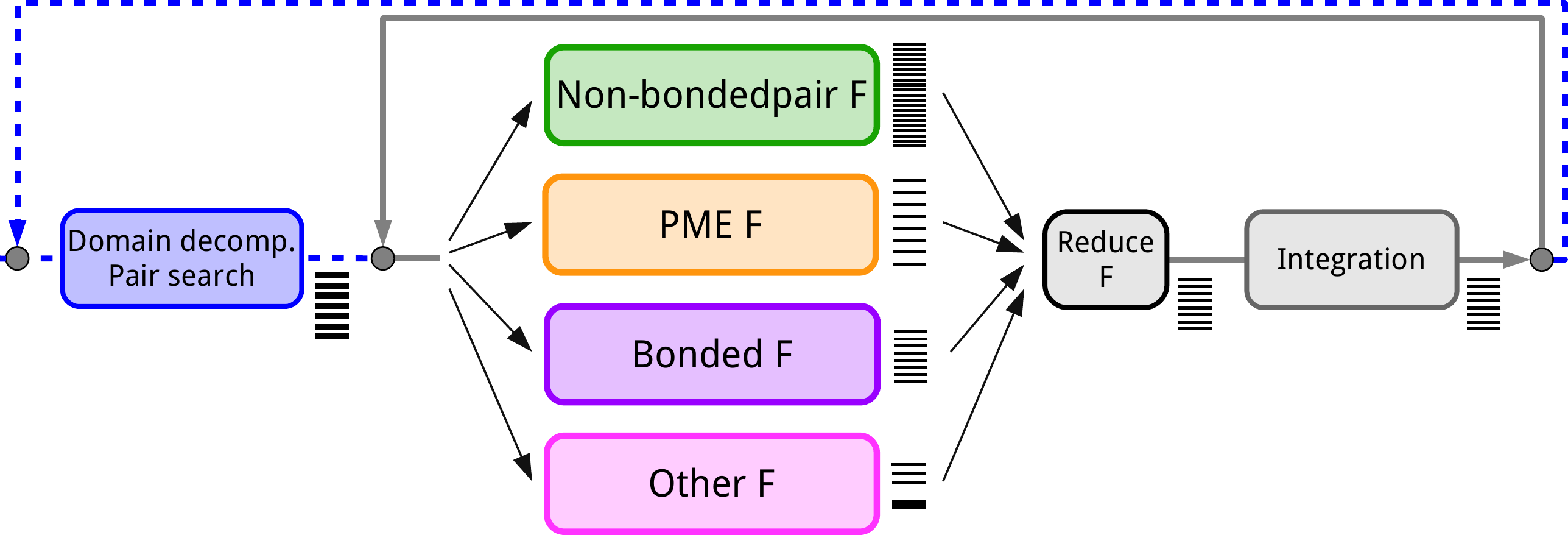}
\caption{Structure of an MD step. There is concurrency available for parallelization both 
    for different force terms and within tasks (horizontal bars).
    The inner (grey) loop only computes forces and integrates, while the less
    frequent outer loop (blue dashed) involves tasks to decompose the problem.
}
\label{fig:md_concurrent_tasks}
\end{figure}

The total force on each particle involves multiple terms: non-bonded pair interactions
(typically Lennard-Jones and electrostatics), bonded interactions (e.g.\ bonds, angles, torsions), and
possibly terms like restraints or external forces.
Given particle coordinates, each term can be computed independently (\figref{fig:md_concurrent_tasks}). 

While there are good examples of MD applications with gigantic systems\cite{Perilla2017},
the most common approach is to keep the simulation size fixed and small to improve absolute performance. Improving the time-to-solution thus requires strong scaling. Historically, virtually all time was spent computing forces, which made it straightforward to parallelize, but well-optimized MD engines now routinely achieve step iteration rates at or below a millisecond \cite{Hess2008,Pronk2013,Abraham2015,Lee2018b}.
Thus, the MD problem is increasingly becoming latency-sensitive where synchronization, bandwidth
and latency both within nodes and over the network are becoming major challenges for homogeneous
as well as heterogeneous parallelization due to Amdahl's law. This has partly been compensated for
by increasing density of HPC resources where jobs rely more on intra-node communication than
the inter-node network. This in turn has enabled a shift from coarse parallelism using MPI and domain decomposition to finer-grained concurrency within force calculation tasks,
which is better suited for multicore and accelerator architectures. 

Dense multi-GPU servers often make it possible for simulations to remain on a single node,
but as the performance has improved the previous fast intra-node communication has become the new 
bottleneck compared to the fast synchronization within a single accelerator or CPU. As a side-effect,
simulations that a decade ago required extreme HPC resources are now routinely performed on single nodes
(often with amazingly cost-efficient consumer hardware). This has commoditized MD simulations,
but to explore biological events we depend on advancing absolute performance such
that individual simulations cover dynamics in the hundreds of microseconds, which can then be
combined with ensemble-level parallelism to sample multi-millisecond processes.

In the mid 2000s, processors hit a frequency wall and the increases in transistor count were
instead used for more functional units, leading to multi- and manycore designs.
Specialization has enabled improvements such as wider and more feature-rich SIMD units as well as
application-specific instructions, and recent GPUs have also brought compute-oriented
application-specific acceleration features.
Compute units however need to be fed data,
but memory subsystems and interconnects have not showed similar improvement. Instead, there has been an increasing discrepancy between the speed of computation and data movement.
The arithmetic intensity per memory bandwidth required to fully utilize compute units 
has increased threefold for CPUs in the last decade, and nearly tenfold for GPUs\cite{Rupp2019}. In particular, most accelerators still rely on communication over the slowly evolving PCIe bus while their peak FLOP rate
has increased five times and GROMACS compute kernel performance grew by up to an order of magnitude\cite{Kutzner2019}.
This imbalance has made heterogeneous acceleration and overlapping compute and communication increasingly difficult. This is partly being addressed through tighter host--accelerator integration with NVIDIA NVLink among the first (other technologies include Intel CXL or AMD Infinity Fabric),
and this trend is likely to continue.

MD as a field was established in an era where the all-important goal was to save  
arithmetic operations, which is even reflected in functional forms such as the Lennard-Jones potential
(the power-12 repulsion is used as the square of the power-6 dispersion instead of an expensive exponential). 
However, algorithm design and parallelization is shifting from saving FLOPS to efficient data layout,
reducing and optimizing data movement, overlapping communication and computation, or simply recomputing data
instead of storing and reloading. This is shifting burden of extracting performance to the software,
including authors of compilers, libraries and applications, but it pays off with significant performance
improvements, and the resulting surplus of FLOPS suddenly available will enable the
introduction of more accurate functional forms without excessive cost.

%%%%%%%%%%%%%%%%%%%%%%%%%%%%%%%%%%%%%%%%%%%%%%%%%%%%%%%%%%%%%%%%%%%%%%%%%%%%%%%%%%%%%%%%%%
\section{Parallelization of MD in GROMACS}\label{sec:parallelization-background}
\subsection{The structure of the MD algorithm}
The force terms computed in MD are independent and expose 
task parallelism within each MD step. Force tasks typically only depend on positions from
the previous step and other constant data, although domain decomposition (DD) introduces an
additional dependency on the communication of particle coordinates from other nodes. The per-step concurrency in computing forces (\figref{fig:md_concurrent_tasks}) is a central aspect in 
offload-based parallel implementations. 
The reduction to sum forces for integration acts as a barrier; only when new coordinates become available can the next iteration start. The domain decomposition algorithm on the other hand exposes coarse-grained data parallelism through
a spatial decomposition of the particle system. Within each domain, finer-grained data-parallelism
is also available (in particular in non-bonded pair interactions), but to improve absolute performance it is the total step iteration rate in this high-level flowchart that has to be reduced to the
order of \SI{100}{\micro s}.

\subsection{Multi-level parallelism}\label{sec:multi_level_parallelism}
Modern hardware exposes multiple levels of parallelism characterized by the type and speed of
data access and communication between compute units.
Hierarchical memory as well as intra- and inter-node interconnects
facilitate handling data close to compute units.
Targeting each level of parallelism has been increasingly important on petascale architectures,
and GROMACS does so to improve performance\cite{Pall2014,Abraham2015}.

On the lowest level, SIMD units of CPUs offer fine-grained data-parallel execution.
Similarly, modern GPUs rely on SIMD-like execution called
SIMT (single instruction, multiple thread) where a group of threads execute in lockstep
(width typically 32--64). CPUs have multiple cores, frequently with multiple hardware threads per core. 
Similarly, GPUs contain groups of execution units (multiprocessors/compute units),
but unlike on CPUs distributing work across these cannot be controlled directly, which poses load balancing challenges.
On the node level, multiple CPUs communicate through the system bus. Accelerators are attached to the host CPU or other GPUs
using a dedicated bus. These CPU--GPU and GPU--GPU buses 
add complexity in heterogeneous systems and, together with the separate global memory,
represent some of the main challenges in a heterogeneous setup.
Finally, the network is essentially a third-level bus for inter-node communication. 
An important concern is not only fast access on each level, but also the non-uniform
memory access (NUMA): moving data between compute units has non-uniform cost. 
This also applies to intra-node buses as each accelerator is typically connected
only to one NUMA domain, not to mention inter-node interconnects where the topology
can have large impact on communication latency.

The original GROMACS approach was largely focused on 
efficient use of low- to medium-scale resources, in particular commodity hardware, through highly
tuned assembly (and later SIMD) kernels. The original MPI- (and PVM) based scaling was 
less impressive, but in version 4.0\cite{Hess2008} this was replaced with a state-of-the-art
neutral-territory domain-decomposition\cite{VanDerSpoel2005} combined with fully flexible
3D dynamic load balancing (DLB) of triclinic domains. This is combined with a high-level task
decomposition that dedicates a subset of MPI ranks to long-range PME electrostatics to
reduce the cost of collective communication required by the 3D FFTs, which means
multiple-program, multiple-data (MPMD) parallelization. Domain decomposition was initially also
used for intra-node parallelism using MPI as well as our own thread-based MPI library implementation\cite{Pronk2015}.
Since the DD algorithm ensures data locality, this has been a
surprisingly good fit to NUMA architectures, but it comes with challenges related to
exposing finer-grain parallelism across cores and limits the ability to make use of efficient
data-exchange with shared caches. Algorithmic limitations (minimum domain size) also restrict
the amount of parallelism that can be exposed in this manner. While the design had served well,
significant extensions were required in order to target manycore and heterogeneous GPU-accelerated
architectures.

The multilevel heterogeneous parallelization was born from a redesign that extended
the parallelization to separately target each level of hardware parallelism, first 
introduced in version 4.6\cite{Abraham2015}.
New algorithms and programming models have been adopted to expose parallelism with finer granularity.
Our first major change was to redesign the pair-interaction calculation to provide a flexible and future-proof algorithm with accuracy-based settings and load balancing capabilities,
which can target either wide SIMD or GPU architectures. On the CPU front, SIMD parallelism is used for
most major time-consuming parts of the code. This was
necessitated by Amdahl’s law: as the performance of non-bonded kernels and PME improved, previously
insignificant components such as integration turned into new bottlenecks. This was made fully portable
by the introduction of the GROMACS SIMD abstraction layer, which started as the
replacement of raw assembly with intrinsics and now supports a
range of CPU architectures using 14 different SIMD instruction sets\cite{GROMACS2020}, with 
additional ones in development.

To utilize both CPUs and GPUs, intra-node parallelization was extended with
an accelerator offload layer and multithreading. The offload layer schedules GPU tasks and data
movement to ensure concurrent CPU--GPU execution and it has evolved as more offload abilities were added.
OpenMP multithreading was first introduced to improve PME scaling\cite{Schulz2009},
and later extended to the entire MD engine. To allow assembling larger units of
computation for GPUs, we increased the size of MPI tasks and
have them run across multiple cores instead of dispatching work from a large number of MPI ranks per node.
This avoids bottlenecks in scheduling and execution of small GPU tasks. Multithreading algorithms
have also gone through several generations with a focus on data-locality, cache-optimizations,
and load balancing, improving scalability to larger thread counts. 
Hardware topology detection based on the hwloc\cite{Broquedis2010} library is used
to guide automated thread affinity setting, and NUMA considerations are taken into account when placing threads.

\begin{figure}
\center
\includegraphics[width=0.95\columnwidth]{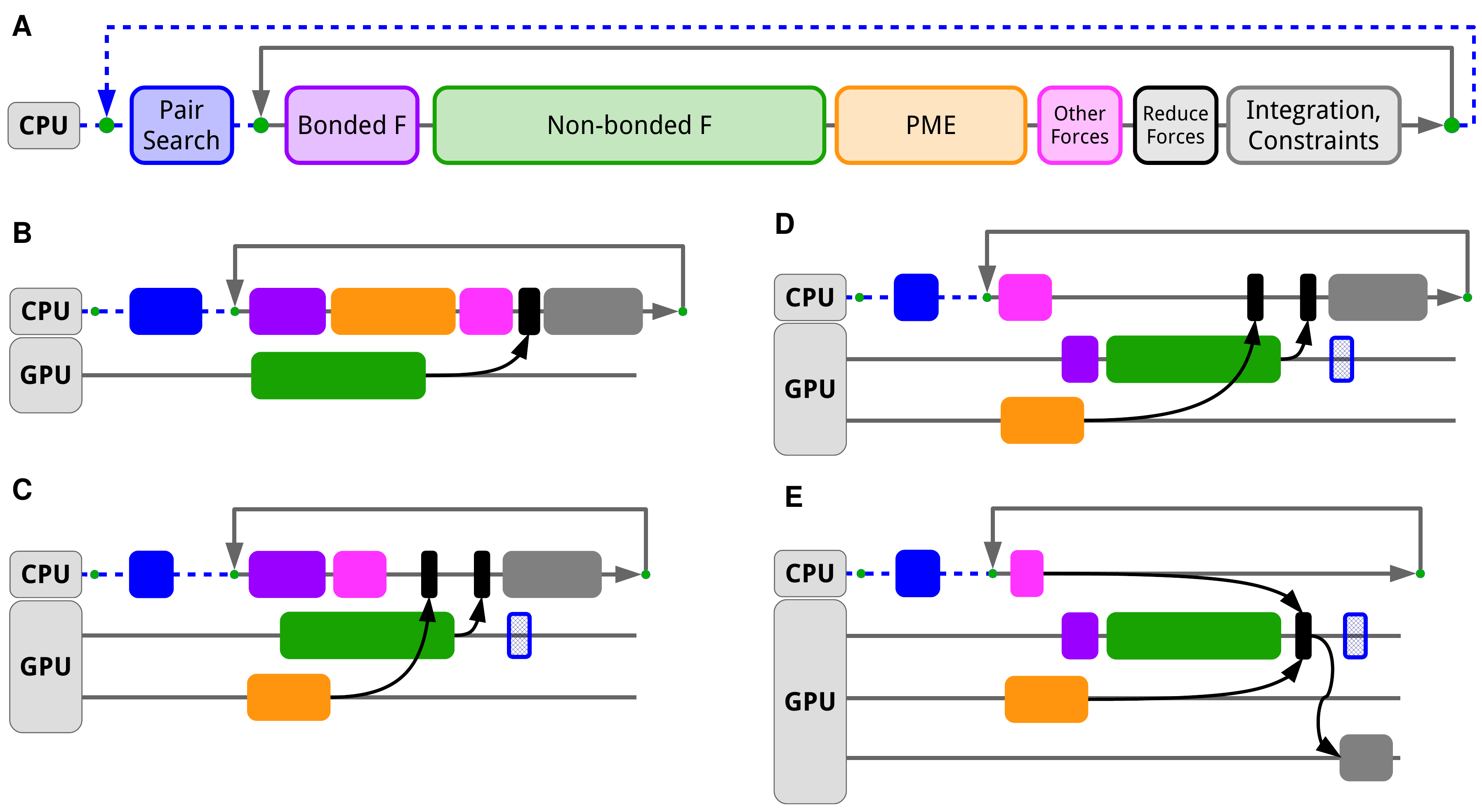}
\caption{
Execution flows of (A)
single-CPU and flavors of the CPU--GPU heterogeneous offload
designs. Incremental task offloading is shown for (B) short-range non-bonded interactions (green),
(C) PME (orange) and dynamic list pruning (blue cross-hatch), (D) bonded interactions (purple), and (E) integration/constraints (gray).
With asynchronous offload, force reduction requires resolving dependencies. The explicit ones are enforced through synchronization (CPU) or
asynchronous events (GPU) illustrated by black arrows. Implicit synchronization is fulfilled
by the sequential dependencies between tasks on the CPU timeline,
as well as those on the same GPU timeline (corresponding to in-order streams).
}
\label{fig:heterogeneous-offload-designs}
\end{figure}

For single-node CPU-only parallelism, execution is still done with sequentially dependent tasks (\figref{fig:heterogeneous-offload-designs}A), which allows relying on implicit dependencies and sharing
output across force calculations. 
Expressing concurrency (\figref{fig:md_concurrent_tasks}) to allow parallel execution over
multiple cores and GPU has required explicitly expressing dependencies.
The new design uses multi-threading and heterogeneous extensions for handling force accumulation and reduction.
On the CPU, per-thread force accumulation buffers are used with cache-efficient sparse reduction
instead of atomic operations. This is important e.g.\
for bonded interactions where a thread typically only contributes forces to a small
fraction of particles in a domain. When combined with accelerators, force tasks can be assigned
to either CPU or GPU, with additional remote force contributions received over MPI.
With forces distributed in CPU and GPU memories, we use a new reduction tree to combine all
contributions. 
Explicit dependencies of this reduction for the single GPU case are indicated by black arrows in \figref{fig:heterogeneous-offload-designs}.
Fulfilling dependencies may require CPU--GPU transfers to the compute unit that does the reduction,
and the heterogeneous schedule is optimized to ensure these overlap with computation
(\figref{fig:heterogeneous-offload-designs} panels C and D).

%%%%%%%%%%%%%%%%%%%%%%%%%%%%%%%%%%%%%%%%%%%%%%%%%%%%%%%%%%%%%%%%%%%%%%%%%%%%%%%%%%%%%%%%%%
\section{Heterogeneous parallelization}\label{sec:evolution-of-parallel-design}
Asynchronous offloading in GROMACS is implemented using either CUDA or OpenCL APIs,
and has two main functionalities: explicit control of CPU–GPU data movement as well as
asynchronous scheduling of concurrent task execution and synchronization. This design aims to maximize CPU--GPU
execution overlap, reduce the number of transfers by moving data early, 
keeping data on the accelerator as long as possible, ensuring transfer is
overlapping with computation, and optimize task scheduling for the critical path to reduce the time per step.

\subsection{Offloading force computation}
\begin{figure*}
\center
\includegraphics[width=0.75\textwidth]{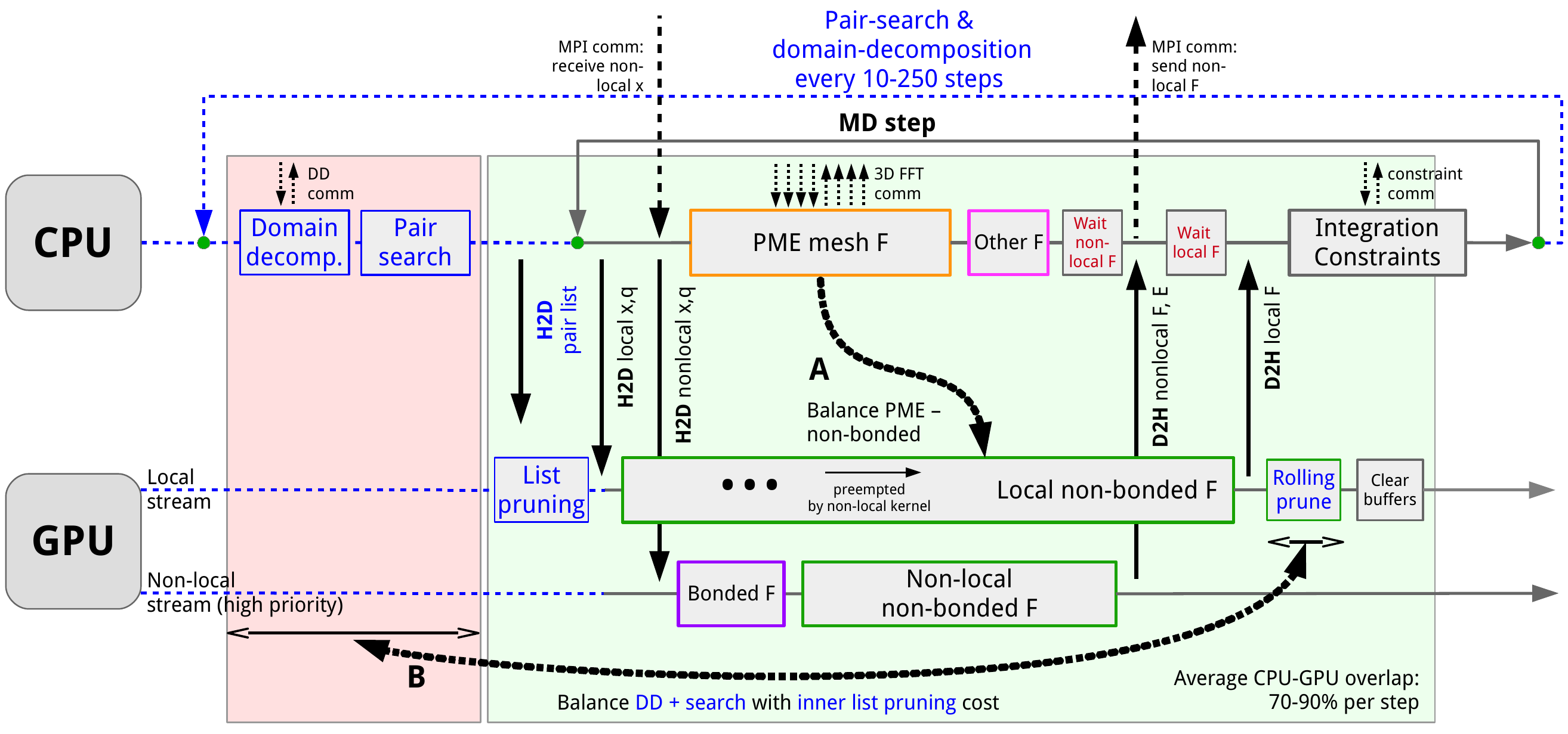}
\caption{Multi-domain heterogeneous control and data flow with short-range non-bonded and
bonded tasks offloaded. Horizontal lines indicate CPU/GPU timelines with inner MD steps (grey) 
    and pair-search/DD (dashed blue). Data transfers are indicated by
    vertical arrows (solid for CPU--GPU, dashed for MPI; H2D is host-to-device, D2H device-to-host).
    The area enclosed in green
    is concurrent CPU--GPU execution, while the red indicates no overlap (pair search and DD).
    Task load balancing is used to increase CPU--GPU overlap (dotted arrows) by shifting work
    between PME and short-range non-bonded tasks (A) and balancing CPU-based pair search/DD with list pruning (B).
}
\label{fig:control-flow-multi}
\end{figure*}

GROMACS initially chose to offload the non-bonded pair interactions to the GPU,
while overlapping with PME and bonded interactions being evaluated on the CPU
(\figref{fig:heterogeneous-offload-designs}B).
While this approach requires CPU resources,
it has the advantage of supporting domain decomposition as well as all functionality,
since any special algorithm can execute on the CPU\cite{Pall2014,Kutzner2015}.
When combined with DD,
interactions with particles not local to the domain depend on halo exchange. This
is handled by splitting non-bonded work into two kernels: one for local-only interactions and one
for interactions that involve non-local particles. Based on co-design with NVIDIA, stream
priority bits were introduced in the GPU hardware and exposed in CUDA. This made it possible to launch
non-local work in a high priority stream to preempt the local kernel and return
remote forces early, while the local kernel execution can overlap with communication. Currently
only a single 
priority bit is available, but increasing this should facilitate additional offloading; this is a less complex solution than e.g.\ a persistent kernel with dynamic workload-switching. 
The intra-node load balancing, together with control and data flow of the heterogeneous
setup with with short-range force offload (non-bonded and bonded) is illustrated in \figref{fig:control-flow-multi}.

The gradual shift in CPU--GPU performance balance in heterogeneous systems\cite{Kutzner2019} brought the need
for offloading further force tasks to avoid the CPU becoming a bottleneck or, from a
cost perspective, not needing expensive CPUs. Consequently, we added offload of PME long-range electrostatics both in CUDA and OpenCL.
PME is offloaded to a separate stream (\figref{fig:heterogeneous-offload-designs}C)
to allow overlap with short-range interactions. The main challenge arises from the two 3D FFTs required.
Simulations rely on small grids (typical dimensions $32$--$256$),
which has not been an optimization target in GPU FFT libraries, so scaling FFT to multiple GPUs would often not
provide meaningful benefits. However, we can still allow multi-GPU scaling by reusing our MPMD 
approach and placing the entire PME execution on a specific GPU. We have also developed a 
hybrid PME offload to allow back-offloading the FFTs to the CPU, while the rest of the work is done on the GPU.
This is particularly useful for legacy GPU architectures where FFT performance can be low.
Additionally, it could be beneficial for strong scaling on next-generation 
machines with high-bandwidth CPU--GPU interconnects to allow grid transfer overlap
and exploiting well-optimized parallel CPU 3D FFT implementations.
The last force task to be offloaded was the bonded interactions. Without DD, this is executed on the
same stream as the short-range non-bonded task (\figref{fig:heterogeneous-offload-designs}D).
Both tasks consume the same non-bonded layout-optimized coordinates and share force output buffer.
With DD, the bonded task is scheduled on the nonlocal stream (\figref{fig:control-flow-multi})
as it is often small and not split by locality.

The force offload design inherently requires data transfer to/from the GPU every step,
copying coordinates to, and forces from, the GPU prior to force reduction
(black boxes in \figref{fig:heterogeneous-offload-designs}B--E),
followed by integration on the CPU. With accelerator-heavy systems this can render a
offload-based setup CPU-limited. In addition, GPU compute to PCIe bandwidth is also imbalanced. 
High performance interconnects are not common, and typically used only
for GPU--GPU communication. This disadvantages the offload design as it leaves the GPU
idle for part of each step, although this can partly be compensated for with
pair list pruning described in the algorithms section.
\change
Pipelining 
\stopchange
force computation, transfer and
integration or using intra-domain force decomposition can reduce the CPU bottlenecks\cite{Stone2016}.
However, slow CPU--GPU transfers are harder to address by overlapping since 
computation is faster than data movement.
For this reason our recent efforts have aimed to increasingly eliminate CPU--GPU data movement and rely on direct GPU communication for scaling.

\subsection{Offloading complete MD iterations}
To avoid the data transfer overhead , GROMACS now supports executing the
entire innermost iteration, including integration, on accelerators (\figref{fig:heterogeneous-offload-designs}E).
This can fully remove the CPU from the critical path, and reduces
the number of synchronization events. At the same time, the CPU
is employed for pair search and domain
decomposition (done infrequently), and special algorithms can use the now free CPU resources during the GPU step. In addition to the force tasks performed on the GPU, 
the data ownership for the particle coordinates, velocities and forces is now also moved to the GPU. This allows shifting the previous parallelization trade-off and minimize GPU idle time. The inner MD loop however still supports forces that are computed on the CPU,
and often the cycle of copying the data from/to the GPU and evaluating these
forces on the CPU takes less time than the force tasks assigned to the GPU
(\figref{fig:heterogeneous-offload-designs}E). The CPU can now be considered
a supporting device to evaluate forces not implemented on GPU (e.g.\ CMAP corrections),
or those not well-suited for GPU evaluation (e.g.\ pulling forces acting on a single atom).
This keeps the GPU code base to a minimum and balances the load by assigning a different set of tasks to the GPU depending on the
simulation setup and hardware configuration. This is highly beneficial
both for high-throughput and multi-simulation experiments on GPU-dense compute resources
or upgrading old systems with a high-end GPU. At the same time, it also allows making efficient use of
communication directly between GPUs, including dedicated high-bandwidth/low-latency interconnects where available. 
In our most recent implementation, data movement can be automatically routed directly between GPUs instead of
staging communication through CPU memory. When a CUDA-aware MPI library is used, communication operates directly on GPU memory spaces. Our own thread-MPI implementation relies on direct CUDA copies.
Additionally, by exchanging CUDA events, it can use stream synchronization across devices
which allows fully asynchronous communication offload leaving the CPU free from both compute and coordination/wait. The external MPI implementation requires additional CPU--GPU synchronization prior to communication,
but allows the new functionality to be used across
multiple nodes. Much of the GPU--GPU communication, either between short-range tasks or between short-range and PME tasks, is of a halo exchange nature where non-contiguous coordinates and forces are exchanged, which
requires GPU buffer packing and unpacking operations. In particular for this, keeping the outer loop of domain
decomposition and pair search on the CPU turns out to be a clear advantage,
since the index map building is a somewhat complex random access operation, but
once complete the data is moved to the accelerator and reused across multiple simulation time steps.

%%%%%%%%%%%%%%%%%%%%%%%%%%%%%%%%%%%%%%%%%%%%%%%%%%%%%%%%%%%%%%%%%%%%%%%%%%%%%%%%%%%%%%%%%%
\section{Algorithm details}\label{sec:core-algorithms}
\subsection{The cluster pair algorithm}\label{subsec:cluster-pair-alg}
The Verlet list\cite{Verlet1967} and linked cell list\cite{Hockney1974} algorithms for
finding particles in spatial proximity
and constructing lists of short-range neighbors were some of the first algorithms in the
field and are cornerstones of MD. However, while the Verlet list exposes a high degree of 
parallelism its traditional formulation expresses this
in an irregular form, which is ill-suited for SIMD-like architectures.
To reduce the execution imbalance due to varying list lengths, 
padding\cite{Yang2007} or binning\cite{Stone2007} have been used.
However, the community has largely converged on reformulating the problem 
by grouping interactions into fixed size work units instead.\cite{VanMeel2008,Friedrichs2009,Gotz2012,Pall2013,Stone2016,Kobayashi2017}

A common approach is to assign different $i$-particles to each SIMT thread
requiring a separate $j$ particle loaded for each pair interaction.
This leads to memory access divergence which becomes a bottleneck in SIMD-style
implementations, even with arithmetically intensive interactions.\cite{Pennycook2013}
Sorting particles to increase spatial locality for better caching improves
performance\cite{Gonnet2007,Anderson2008,Eastman2009,Welling2011},
but the inherently scattered access is still inefficient.
Early efforts used GPU textures \cite{Stone2007,Bailey2015} to improve data reuse,
but this is hard to control as the effectiveness
depends on the size of the $j$-lists relative to cache.

\begin{figure*}
\center
\includegraphics[width=0.75\textwidth]{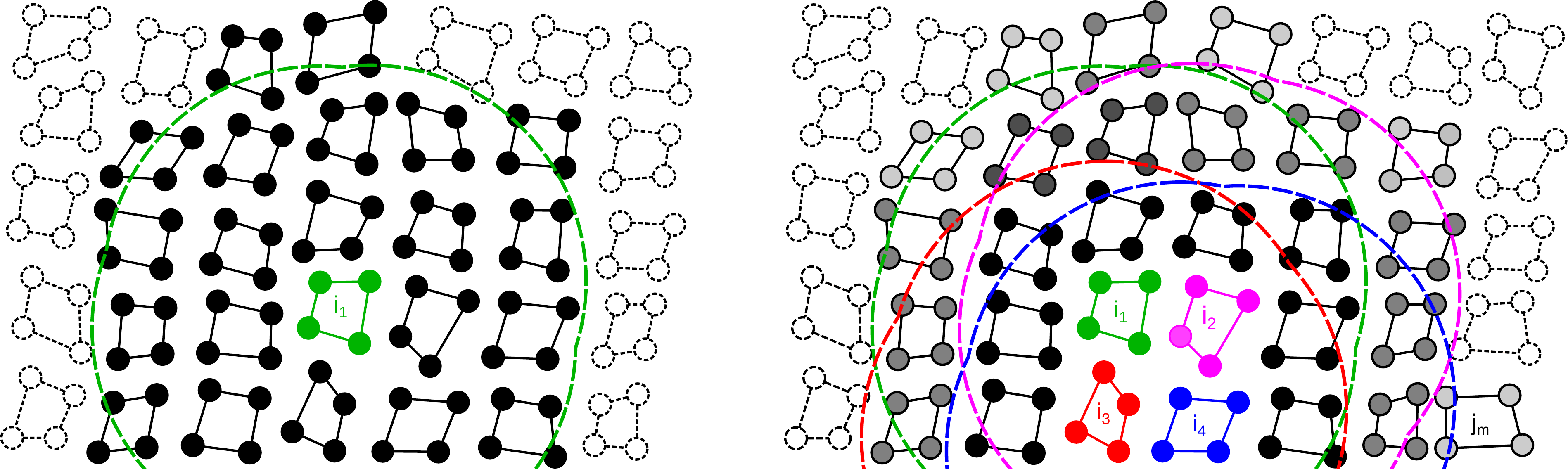}
\caption{Cluster pair setups with four particles ($N=4$, $M=4$).
    Left: CPU/SIMD-centric setup. All clusters with solid lines are included in the pair list of cluster
    $i_1$ (green). Clusters with filled circles have interactions within the buffered cut-off (green dashed line) of at least one particle in $i_1$,
    while particles in clusters intersected by the buffered cut-off that fall outside of it represent an extra implicit buffer.
    Right: Hierarchical super-clusters on GPUs.
    Clusters $i_1$--$i_4$ (green, magenta, red, and blue)
    are grouped into a super-cluster. Dashed lines
    represent buffered cut-offs of each $i$-cluster. Clusters with any particle in any region will be
    included in the common pair list. Particles of $j$-clusters in the joint list are illustrated 
    by discs filled in black to gray; black indicates clusters that interact with all four $i$-clusters,
    while lighter gray shading indicates a cluster only interacts with 1--3 $i$-cluster(s), e.g.\ $j_m$
    only with $i_{4}$.
}
\label{fig:cluster-setup}
\end{figure*}

Given the need for increasing the arithmetic-to-memory operation ratio,
we formulated an algorithm that regularizes the problem and increases
data reuse. The goal is to load $j$-particle data as efficiently and rarely as possible 
and reuse it for multiple $i$-particles, roughly similar to blocking algorithms in matrix-matrix multiplication.
Our cluster pair algorithm uses a fixed number of particles per cluster, and pairs of such 
clusters rather than individual particles are the unit of computing short-range interactions.
Hence, we compute interactions between $i$-clusters of $N$ particles and
a list of $j$-clusters each of $M$ particles. $M$ is adjusted to the SIMD width while $N$ allows
balancing data reuse with register usage. In addition to a data layout that allows efficient access,
and that $N \times M$ interactions are calculated for every load/store, the algorithm is easy to adapt to
new architectures or SIMD widths.
Since the algorithmic efficiency will be
higher for smaller clusters, we can also place
two sets of the $N$ $i$-cluster particles in a wide SIMD register of length $2M$, which our SIMD
layer supports on all hardware where sub-register load/store operations are efficient. 
The clusters and pair list are obtained during search using a regular grid in $x$ and $y$ dimensions but
binning the resulting $z$ columns into cells with fixed number of particles (in contrast to fixed-size cells) 
that define our clusters (\figref{fig:cluster-setup}, left). The irregular 3D grid is then used to build the
cluster pair list\cite{Pall2013}.

Following the approach used for CPU SIMD, choosing $M$ to match the GPU execution width may seem suitable.
However, as GPUs typically have large execution width, the resulting cluster size would greatly
increase the fraction of zero interactions evaluated. The raw FLOP-rate would be high, but efficiency low.
To avoid this, our GPU algorithm calculates
interactions between all pairs of an $i-j$ cluster pair instead of assigning the same $i$- and different
$j$-particles to each GPU hardware thread. Hence, we adjust $N \times M$ to the GPU execution width.
However, evaluating all $N \times M$ interactions of a cluster pair in parallel would
eliminate the $i$-particle data reuse. The arithmetic intensity required to saturate modern processors is
quite similar across the board\cite{Barba2013,Rupp2019}, so restoring data reuse 
is imperative. We achieve this by introducing a \emph{super-cluster} grouping (\figref{fig:cluster-setup}).
A \emph{joint pair list} is built for the super-cluster as the union of $j$-clusters that
fall in the interaction sphere of any $i$-cluster. This improves arithmetic saturation at the cost of some
pairs in the list not containing interacting particles, since all $j$-clusters are not shared.
To minimize this overhead, the super-clusters are kept as compact as possible, and the 
search is optimized to obtain close to cubic cluster geometry -- we use an eight-way grouping obtained from a
$2 \times 2 \times 2$ cell grouping on the search grid. Even so, the joint $j$-list would lead to
substantial overhead if interactions were computed with all clusters, similarly to the challenge with large regular tiling. We avoid this elegantly by skipping cluster pairs with no interacting particles
based on an interaction bitmask stored in the the pair list.
This is illustrated in \figref{fig:cluster-setup} where lighter-shaded $j$-clusters do not interact
with some of the $i$-clusters; e.g.\ $j_m$ can be skipped for $i_1$--$i_3$.
As $M \times N$ is adjusted to match the execution width,
the interaction masks allow efficient divergence-free skipping of entire cluster pairs.
Our organization of pair-interaction calculation is similar to that used by others \cite{Friedrichs2009,Salomon-Ferrer2013}, with the key difference that those approaches rely on larger fixed size tiles and use other techniques to reduce the impact of large grouping.

The interaction mask describes a $j-i$ relationship, swapping the order of the standard
$i-j$ formulation. Consequently, the loop order is also swapped and our GPU implementation uses an outer
loop over the joint $j$-cluster list and inner loop over the eight $i$-clusters.
This has two main benefits: First, 8 bits per $j$-cluster is sufficient to encode the interaction mask,
instead of needing variable-length structure. Second, the force reduction becomes more efficient. 
Since we utilize Newton's third law to only calculate interactions once, we need
to reduce forces both for $i$- and $j$-particles. At the end of an outer $j$ iteration, all interactions of
the $j$-particles loaded will have been computed and the results can be reduced and stored.
At the same time, accumulating the $i$-particle partial forces requires little memory ($8 \times 8$ forces)
and can be done in registers. Self-exclusions are handled in the interaction kernels while force-field
exclusions are stored in the list with $j$-clusters as bitmasks and enforced simultaneously
with the interaction cut-off, just as the interaction masks\cite{Pall2013}.
In our typical target systems, on average approximately four
of the eight $i$-clusters contain interactions with $j$ particles. 
Hence, about half of the inner loop checks result in skips, and
we have observed these to cost 8--12\%, which is a rough
estimate of the super-cluster overhead. In comparison, the 
normal cut-off check has at most 5--10\% cost in the CUDA implementation. 

For PME simulations, we calculate the real-space Ewald correction term in the
kernel. On early GPU architectures (and CPUs with low FMA FLOPS) tabulated $F\cdot r$ is most efficient.
On all recent architectures,
we have instead developed an analytical function approximation of the correction
force. This yields better performance as it relies on FMA arithmetics
despite the $>15\%$ increase in kernel instruction count.

Multiprocessor-level parallelism is provided by assigning each thread block a list
element that computes interactions of all particles in a super-cluster.
To avoid conditionals, separate kernels are used for different combinations
of electrostatics and Lennard-Jones interactions and cut-offs, and whether energy is required or not. 
The cluster algorithm has been implemented both in CUDA and OpenCL and tuned for multiple
GPU architectures.
On NVIDIA and recent AMD GPUs with 32-wide execution we use an $8 \times 4$ cluster setup,
for 64-wide execution on AMD $8 \times 8$, and on Intel hardware with 8-wide
execution a $4 \times 2$ setup, all with 8-way super-clustering.

\subsection{Algorithmic work efficiency and pair-list buffers}\label{subsec:work-efficiency}
\begin{figure}
\center
\includegraphics[width=0.65\columnwidth]{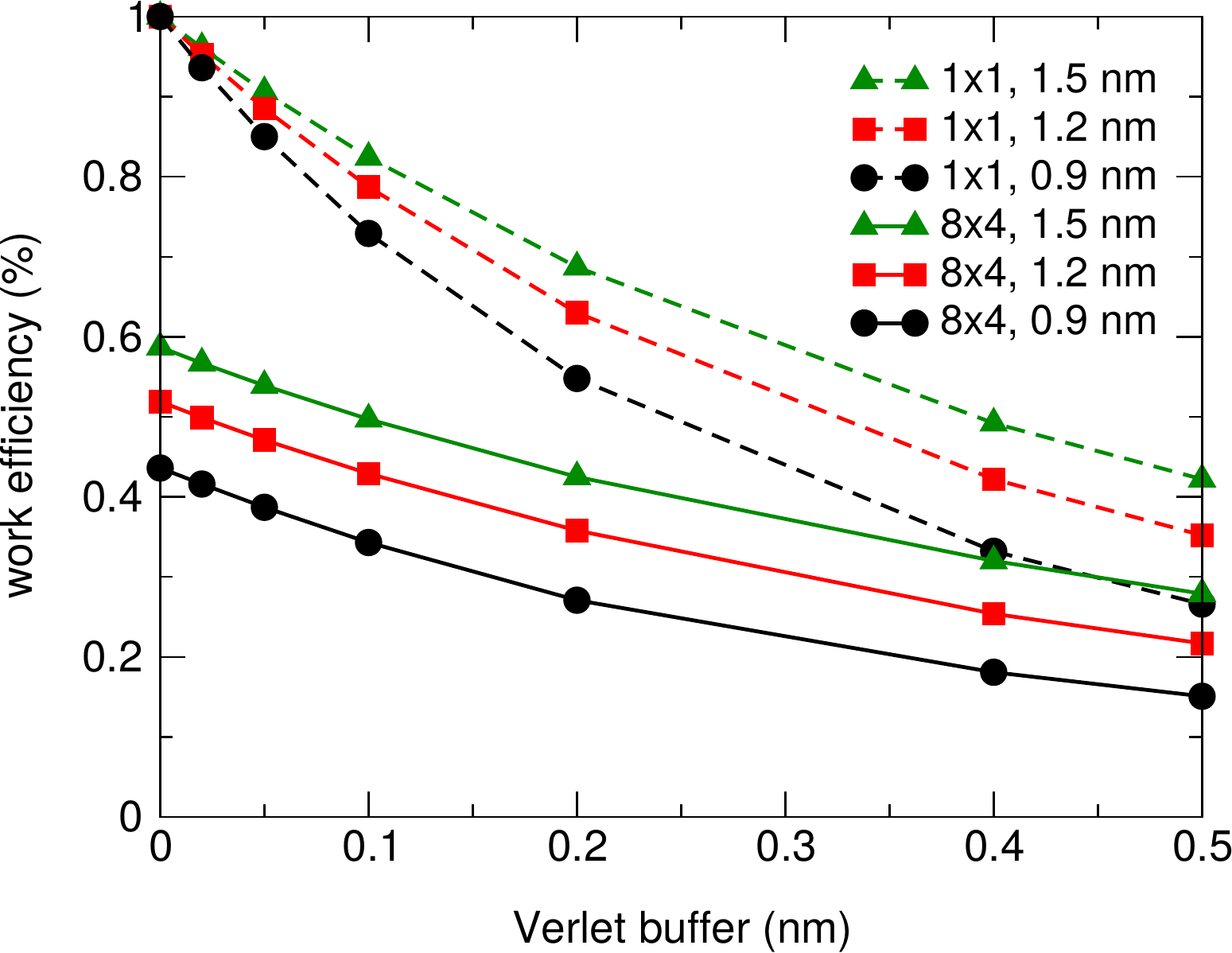}
\caption{
    Algorithmic work-efficiency of the particle ($1 \times 1$) and cluster ($8 \times 4$) Verlet approaches,
    defined as the fraction of interactions calculated that are within
    the actual cut-off, shown as function of buffer size for three common
    cut-off distances. 
    The trade-off is that larger cluster sizes enable greater
    {\em computational} efficiency, and increased buffers enable longer reuse of the pair list.
}
\label{fig:work-efficiency}
\end{figure}

The cluster algorithm trades computing interactions known to evaluate to zero for
improved execution efficiency on SIMD-style architectures. 
To quantify the amount of additional work,
we calculate the parallel work efficiency of the algorithm as
fraction of non-zero interactions evaluated, It is worth noting that this metric is $\leq 1$ even for
the standard Verlet algorithm
as any finite buffer contains non-interacting particles (\figref{fig:work-efficiency}).
In the cluster algorithm this is 
augmented with particles outside the buffered sphere, but where 
another particle in the cluster falls inside it (\figref{fig:cluster-setup}).
The work-efficiency depends on the cut-off, buffer size,
and geometry/size of the clusters that is optimized during search. The cost of this search is the reason why absolute
performance still benefits from larger buffers, just as
kernel execution efficiency benefits from cluster size.

The improved computational efficiency offsets the algorithmic work-efficiency
trade-off for all modern architectures\cite{Pall2013}.
Moreover, particles in the pair list that fall outside
the buffered cut-off can be made use of:
these provide an extra \emph{implicit buffer} that allows using a
shorter explicit buffer when evaluating pair interactions
while maintaining the accuracy of the algorithm. One example
of a cluster contributing to the implicit buffer is illustrated in
\figref{fig:cluster-setup}. Making use of this implicit buffer
increases the practical parallel efficiency relative to the standard particle-based algorithm.
With a box of SPC/E water where a $\SI{0.9}{nm}$ cut-off pair list is reconstructed every 40 steps,
the $1 \times 1$ setup requires a buffer of
$\SI{0.218}{nm}$ to reach the same error tolerance as the 
$8 \times 4$ cluster setup achieves with a $\SI{0.105}{nm}$ buffer.

We believe this is a striking example of the importance of moving
to tolerance-based settings instead of rule-of-thumb or heuristics
to control accuracy.
All algorithms in a simulation affect the accuracy of the final results,
and while the acceptable error varies greatly between problems, it
will be dominated by the worst part of the algorithm -- there is little
benefit from evaluating only some parts more accurately.
To control the effect of missing pair interactions close to the cut-off,
our implementation estimates the Verlet buffer for a given upper bound for the error in the energy.
The estimate is based on the particle masses, temperature, pair interaction functions and constraints,
also taking into account the cluster setup and its implicit buffering \cite{Pall2013}.
Since first introduced, we have refined the buffer estimate to account for
constrained atoms rotating around the atom they are constrained to rather than moving linearly,
which allows tighter estimates for long list lifetimes.
The upper bound for the maximum drift can be provided by the user as a tolerance setting in the simulation input.
We use $\SI{0.005}{kJ/mol/ps}$ per atom as default, but the tolerance % 2 kt/ns
and hence drift can be arbitrarily small. For the default setting,
the implicit buffer turns out to be sufficient for a water system or solvated biomolecule with PME electrostatics and $\SI{20}{\femto \second}$ pair-list update intervals, and
no extra explicit buffer is thus required in this case. The {\em actual}
energy drift caused by these settings is $\SI{0.0001}{kJ/mol/ps}$ per atom, 
a factor of 5 smaller than the upper bound. For comparison, typical constraint algorithms
result in drifts around \SI{0.0002}{kJ/mol/ps} per atom, so being significantly
more conservative than this will usually not improve the overall error in a simulation.

We see several advantages to this approach, and would argue the field in general should
move to requested tolerances instead of heuristic settings. First, the user can set
a single parameter that is easier to reason about and that will be valid across
systems and temperatures. Second, it will enable innovation in new algorithms that maintain accuracy (instead of performance improvements becoming a race towards the least accurate implementation).
Finally, since other parameters
can be optimized freely for the input and run conditions, we can make use if this 
for advanced load balancing to
safely deviate from classical setups by relying on maintaining accuracy rather than arbitrary settings as described below.

\subsection{Non-bonded pair interaction kernel throughput}
\begin{figure}
\center
\includegraphics[width=0.7\columnwidth]{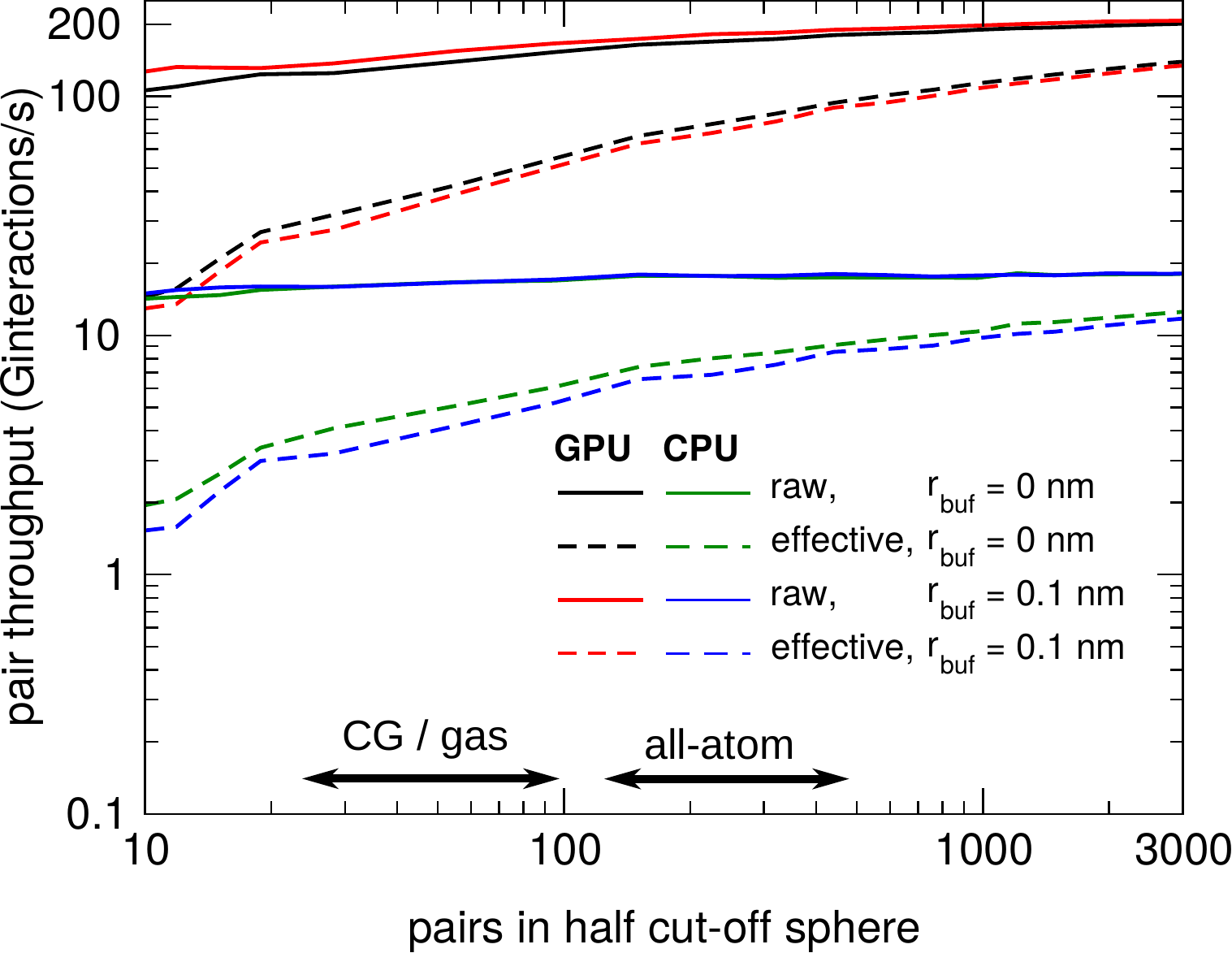}
\caption{
    Non-bonded pair interaction throughput of CUDA GPU and AVX512 CPU kernels as a function of particles in the half cut-off sphere.
    The raw throughput includes zero interactions,
    while effective throughput only counts non-zero interactions. Pair ranges typical to all-atom and coarse-grained/gas simulations are indicated. Measurements were done using a 157k particle Lennard-Jones system to minimize input size effects (density $\rho= \SI{26}{\mathrm{nm}^{-3}}$, $\sigma=\SI{0.3345}{nm}$). Hardware: NVIDIA V100 PCIe GPU, Intel Xeon Gold 6148 CPU.
}
\label{fig:pair-kernel-throughput}
\end{figure}

The throughput of the cluster-based pair interaction 
algorithm depends on the number of interactions per particle, hence particle density.
This varies across applications,
from coarse-grained to all-atom bio-molecular systems or liquid-crystal simulations.
%%% Szilard: added per request from R#2
\change
Hence, we investigate the performance of the pair interaction kernels as a function of particle density.
We measure the pair throughput of the nonbonded kernel using a Lennard-Jones system consisting of argon atoms
to facilitate comparison across a wide range of application areas from physical to biological systems.
\stopchange
%%%
The {\em effective} pair throughput (counting only non-zero interactions)
is also influenced by the buffer length and the conditionally-enforced cut-off in the GPU kernels. When comparing CUDA GPU and AVX512 CPU kernels
with same-size clusters (identical work-efficiency), the raw
throughput reaches peak performance already from ${\sim}150$ pairs per particle on the CPU, while the GPU does not saturate until ${\sim}1000$ pairs (\figref{fig:pair-kernel-throughput}).
This is explained by the increasing $i$-particle data reuse with longer $j$-lists.
The effective pair throughput shows a monotonic increase, since more pairs will be
inside the cut-off with more particles in the interaction
range, as expected from \figref{fig:work-efficiency}.
For typical all-atom simulations, the effective GPU kernel
throughput gets close to \SI{100}{Ginteractions/s} while
the corresponding throughput on a 20-core CPU is an order-of-magnitude lower.

\begin{figure}
\includegraphics[width=0.95\columnwidth]{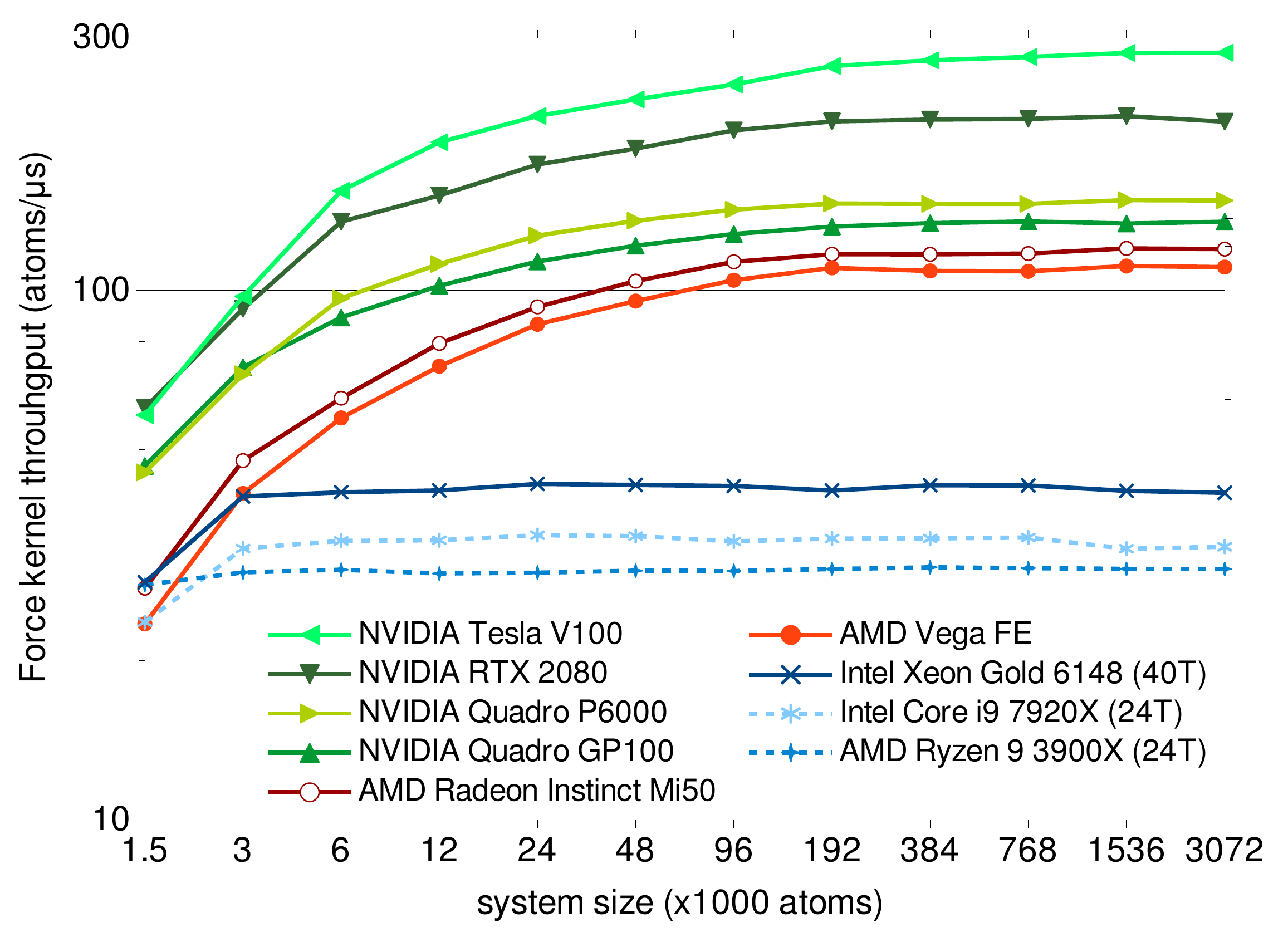}
\caption{
    Force-only pair interaction kernel performance as a function of input size. The throughput indicates how
    CPU kernels have less overhead for small systems, while GPU kernels achieve significantly higher throughput from moderate inputs sizes.
    Multiple generations of consumer and professional CPU and GPU hardware 
    are shown; CUDA kernels are used on NVIDIA GPUs, OpenCL on AMD, AVX2 and AVX512 on the AMD and Intel CPUs, respectively.
    All cores and threads were used for CPUs.
}
\label{fig:nb_kernel_perf_v2}
\end{figure}

To provide good performance for small systems and enable strong scaling,
it is important to achieve high kernel efficiency already at limited particle counts.
To illustrate performance as a function of system size for all-atom systems, 
\figref{fig:nb_kernel_perf_v2} shows
actual pair throughput for SPC/E water systems with \SI{1}{nm} cut-off,
Ewald electrostatics, 40 step search frequency and default tolerances.
%%% Szilard: added per request from R#2
\change
Water typically represents up to 90\% of biomolecular systems (hence it is the particle density of interest)
and therefore nonbonded performance is critical for water. Historically, GROMACS and other codes used
special kernels for water but this no longer works well with SIMD-style architectures.
On the other hand, when some particles have only one type of interaction (hydrogens in 
water typically only have Coulomb interactions), this opens up the possibility for additional optimizations useful on some architectures. 
\stopchange
%%%
The GROMACS CPU kernels achieve peak performance already around 3000 atoms (using up to 40 threads),
and, apart from the largest devices, within 10\% of peak GPU pair throughput is reached around 48k atoms.
GPU throughput is up to seven-fold higher at peak than on CPUs and although it
suffers significantly with smaller inputs (up to five-fold lower than peak),
for all but the very smallest systems the GPU kernels reach higher absolute throughput.
The challenge for small systems is the overhead incurred from 
kernel invocation and other fixed-cost operations.
Pair list balancing also comes at a slight cost,
and while it is effective when there are enough lists to split,
it is limited by the amount of work relative to the size of the GPU.
The sub-10k atom systems simply do not have enough parallelism to 
execute in a balanced manner on the largest GPUs. In contrast, CPU kernels
exhibit a slight decrease for large input due to cache effects.

\subsection{The pair list generation algorithm}\label{subsec:pair-list-build-and-dual-list}
GROMACS uses a fixed pair list lifetime
instead of heuristic updates based on particle displacement,
since the Maxwell-Boltzmann distribution
of velocities means the likelihood of requesting a
pair list update at a step approaches unity as the size
of the system increases.
In addition, the accuracy-based buffer estimate allows the pair search frequency to be picked freely
(it will only influence the buffer size),
unlike the classical approach that requires it to be carefully chosen 
considering simulation settings and run conditions.

We early decided to keep the pair search on the CPU.
Given the complex algorithms involved, our main reason was to ensure parallelization, portability and ease of maintenance, while
getting good performance (reducing GPU idle time) through algorithmic improvements.
The search uses a SIMD-optimized implementation and, to further reduce its cost,
the hierarchical GPU list is initially built using cluster bounding-box distances
avoiding expensive all-to-all particle distance checks\cite{Pall2013}. 
A particle-pair distance based list pruning is carried out on the GPU which
eliminates non-interacting cluster pairs.
This can also further adapt the setup to the hardware execution width by splitting $j$-clusters;
e.g.\ for current Intel GPUs the search produces a $4 \times 4$ cluster setup (\figref{fig:cluster-setup})
which is pruned into $4 \times 2$ for 8-wide execution.
Initial pruning is done the first time the list is processed by a special
version of the kernel (using warp collectives for low extra cost), 
and the pruned list is reused for consecutive MD steps.
Depending on the cut-off and buffer length, pruning reduces the list size by 50--75\%.

\subsection{Dual pair list with dynamic pruning}\label{subsec:dual-pair-list}
Domain decomposition and pair list generation rely on 
irregular data access, and their performance has not improved at the same rate as compute kernels.
Trading their cost for more pair interaction work through increasing
the search frequency has drawbacks.
First, as the buffer increases the  overhead becomes large (\figref{fig:work-efficiency}).
The trade-off is also sensitive to inputs and runtime conditions, with a small optimal window.
In order to address this, we have developed an extension to the cluster algorithm
with a dual pair list setup that uses a longer outer and a short inner list cut-off. 
The outer list is built very infrequently, while frequent pruning steps produce a 
pair list based on the inner cut-off, typically with close to zero explicit buffer.
As pruning operates on regularized particle data layout produced by the pair search,
it comes at a much lower cost (typically $<1\%$ of the total runtime)
than using a long buffer-based pair list in evaluating pair interactions.
This avoids the previous trade-off and reduces the cost of search and DD without force computation overhead.
With GPUs, pruning is done in a rolling fashion scheduled in chunks between force computations of
consecutive MD steps, which allows it to overlap other work such as
CPU integration (\figref{fig:control-flow-multi}).

\subsection{Multi-level load balancing}\label{subsec:multi-level-load-balancing}
Both data and task decomposition contribute to load imbalance
on multiple levels of parallelism.
Sources of data-parallel imbalance include
inherently irregular pair interaction data (varying list sizes),
non-uniform particle density (e.g.\ membrane protein simulations using united-atom lipids)
resulting in non-bonded imbalance across domains, and
non-homogeneously distributed bonded interactions (solvent does not have as many bonds as
a protein). With task-parallelism when using MPMD or GPU offload, task load imbalance
is also a source of imbalance between MPI ranks or CPU and GPU.
Certain algorithmic choices like pair search frequency or electrostatics settings
can shift load between parts of the computation, whether running in parallel or not. 

In particular for small systems it is a challenge to balance work between
the compute units of high-performance GPUs. Especially with irregular work, 
there will be a kernel ``tail'' where only some compute units are active,
which leads to inefficient execution. In addition, with small domains per GPU
with DD there may be too few cluster lists to process in a balanced manner.
To reduce this imbalance and the kernel tail it would be desirable to control
block scheduling, but this is presently not possible on GPUs\cite{Glaser2015}.
Instead, we tune scheduling indirectly through indexing order.
The GPU pair interaction work is reshaped by sorting to avoid long lists being scheduled late.
In addition, when the number of pair lists is too low for efficient execution for given hardware, 
we heuristically split lists to increase the available parallelism.

Kernel tail effects can be also be mitigated by overlapping compute kernels.
This requires enough concurrent work available to fill idle GPU cores
and that it is expressed such that parallel execution is possible.
GPU APIs do not allow fine-grained control of kernel execution and instead
the hardware scheduler decides on an execution strategy.
By using multiple streams/queues with asynchronous event-based dependencies,
our GPU schedule is optimized to maximize the opportunity for kernel overlap.
This reduces the amount of idle GPU resources due to kernel tails
as well as those caused by scheduling gaps during a sequence of short dependent kernels
(e.g.\ the 3D-FFT kernels used in PME).

\begin{figure}
\center
\includegraphics[width=0.7\columnwidth]{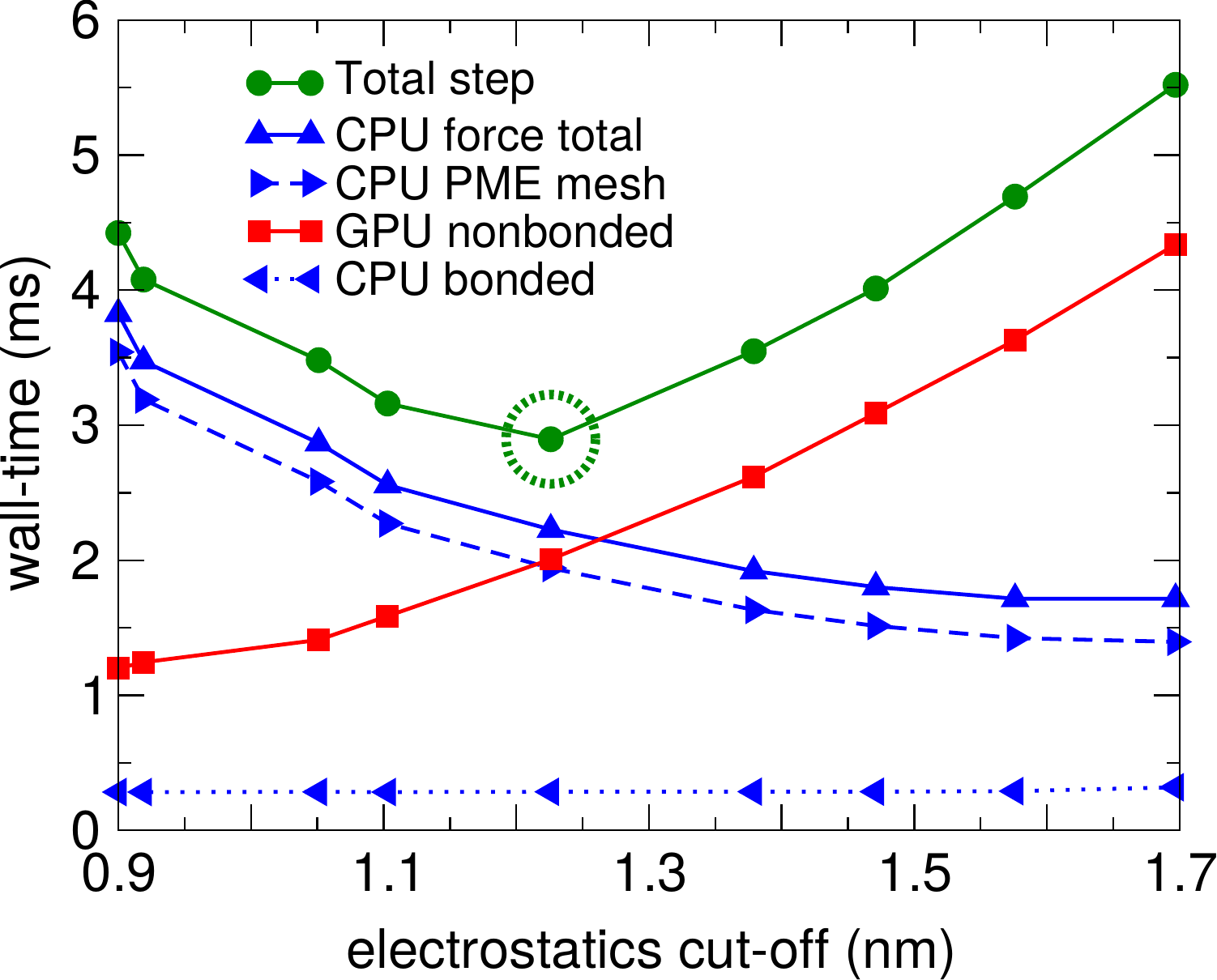}
\caption{
    CPU--GPU load balancing between short- and long-range non-bonded force tasks used.
    The PME load balancing seeks to minimize total wall-time, here at \SI{1.226}{nm} (green dashed circle),
    by increasing the electrostatics direct-space cut-off while also scaling PME grid spacing.
    This shifts load from the CPU PME task (blue dashed) to the non-bonded GPU task (red).
    System: Alcohol dehydrogenase (95k atoms, \SI{0.9}{nm} cut-off, default buffer tolerance).
    Hardware: Intel Core i7-5960X CPU, NVIDIA GTX TITAN GPU.
}
\label{fig:cpu-gpu-lb}
\end{figure}

With PME electrostatics, the split into real and reciprocal space provides
opportunities to rebalance work at constant tolerance by scaling 
the cut-off together with the PME grid spacing\cite{AbrGre2011}. This was first 
introduced as part of or MPMD approach with an external tool\cite{Pronk2013}.
This load balancing was later automated and implemented as an online load balancer\cite{Pall2014}, originally to allow shifting work from the CPU to the GPU. This approach now works remarkably well 
also with dedicated PME ranks both in CPU-only runs and when using multiple GPUs.
The load balancer is automatically triggered at startup
and scans through cut-off and PME grid combinations using
CPU cycle counters to find the highest-performance alternative
as illustrated in \figref{fig:cpu-gpu-lb} 
%%% Szilard: added per request from R#2
\change
using a biomolecular system typical for applications that aim to reduce the system size simulated.
\stopchange
%%%

The load balancer typically converges in a few thousand steps, apart from noisy
environments such as multi-node runs with network contention. Significant efforts have been made to ensure the robustness of the algorithm. It accounts
for measurement noise, avoids cache effects, mitigates interference of CPU/GPU frequency scaling,
and it reduces undesirable interaction with the DD load balancer
(which could change the domain size while the cut-off is scaled).

Nevertheless, load balancing comes with a trade-off in terms of increased communication volume. 
In addition, linearithmic (FFT) or linear (kernel) time-complexity reciprocal-space work is traded for
quadratic time complexity real-space work (\figref{fig:cpu-gpu-lb}). To mitigate waste of energy
we impose a cut-off scaling threshold to avoid increasing GPU load in heavily CPU-bound runs.
The performance gain from PME load balancing depends on the hardware;
with balanced CPU--GPU setups it is up to 25\%, but in
highly imbalanced cases much larger speedups can be observed\cite{Kutzner2015}.

The dual pair list algorithm allows us to avoid most of the drawbacks
of shifting work to direct space, since the list pruning is significantly
cheaper than evaluating interactions. This makes the balancing less sensitive
and easier to use, and where the previous approach saturated around 
pair list update intervals of 50-100 steps, the dual list with pruning
can allow hundreds of steps, which is particularly useful in reducing CPU 
load to maximize GPU utilization in runs that offload the entire inner iteration
(\figref{fig:control-flow-multi}).

Finally, the domain decomposition achieves load balancing 
by resizing spatial domains, thereby redistributing particles between domains and shifting work between MPI ranks. With force offload the use of GPUs is largely
transparent to the code, but extensions to the DLB algorithm were necessary. Support for timing concurrent GPU tasks is limited, in particular in CUDA.
We account for GPU work in DLB through the wall-time the CPU spends waiting for results, labeled accordingly on the CPU timeline in \figref{fig:control-flow-multi}.
However, this can introduce jitter when a GPU is assigned to multiple MPI ranks.
GPUs are not partitioned across MPI ranks, but work 
is scheduled in an undefined order and executed until completion (unless preempted).
Hence, the CPU wait can only be systematically measured on some
of the ranks sharing a GPU while not on others. This leads to spurious imbalance and
to avoid it we redistribute the CPU wall-time spent waiting for the GPU
evenly across the MPI ranks assigned to the same GPU to reflect the average GPU load and eliminate execution order bias.

%%%%%%%%%%%%%%%%%%%%%%%%%%%%%%%%%%%%%%%%%%%%%%%%%%%%%%%%%%%%%%%%%%%%%%%%%%%%%%%%%%%%%%%%%%
\section{Performance benchmarks}\label{sec:performance}

%%
%%% Szilard: added per request from R#2
\change
\paragraph*{Benchmark systems}
%\subsection*{Benchmark systems}
To illustrate the real-world performance of the GROMACS heterogeneous parallelization, we use
a set of benchmark systems representative of typical biomolecular workloads both in term of
size and force-fields.
For single GPU benchmarks we evaluate performance using a small (RNAse) and a medium-sized (GluCl) biomolecular
system, both using the AMBER force field.
To show multi-GPU ensemble simulation throughput we use a medium-sized aquaporin membrane protein with
coupled simulations that employ the Accelerated Weight Histogram (AWH) enhanced sampling algorithm \cite{Lindahl2014}, while 
strong benchmarks use a larger ${\sim 1}$\ million atom satellite tobacco mosaic virus (STMV) system,
still representative of common workloads and challenging for strong scaling.
The latter two benchmark systems use the CHARMM force field and its characteristic settings which
notably yield a different short- to long-range nonbonded workload balance, hence different performance
behavior compared to AMBER-based simulations that use shorter cutoffs.
Further details of the benchmark systems, including input files, can be found in the supplementary
information\cite{SI_GMX_Heterogeneous_JCP_2020}.
\stopchange
%%%

\begin{figure}
\center
\includegraphics[width=0.95\columnwidth]{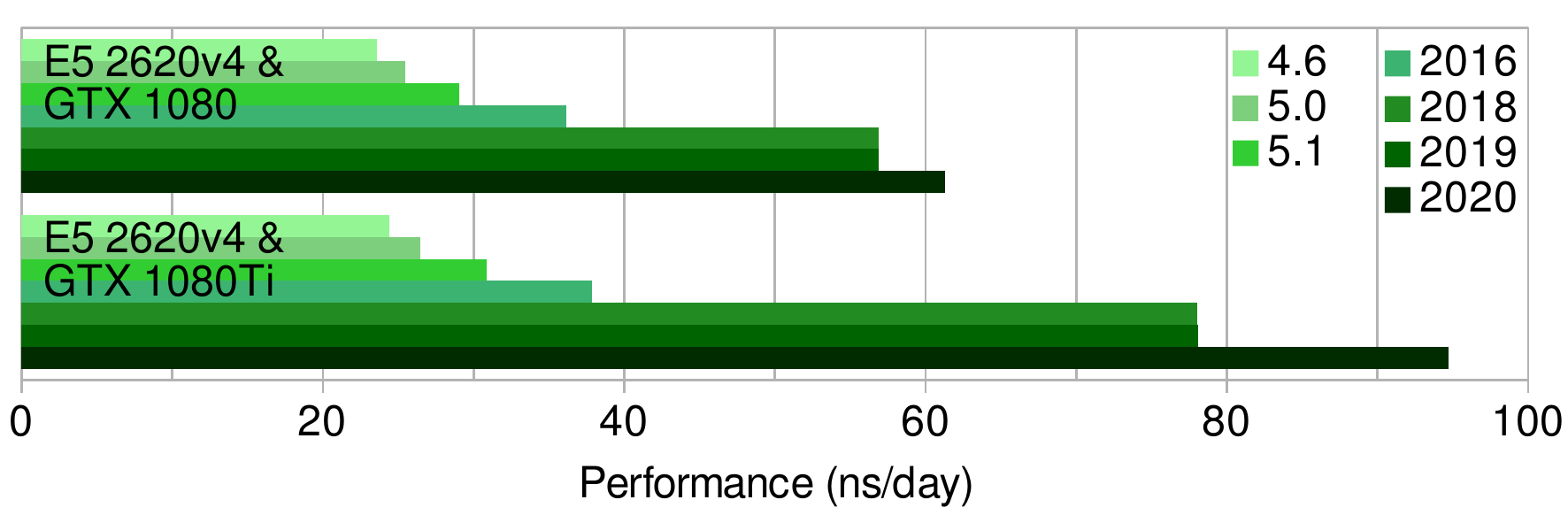}
\caption{
    Performance evolution of GROMACS from the first version with heterogeneous parallelism support on identical hardware.
    Performance is shown for the 142k atom GluCl benchmark
    on two hardware configuration with varying CPU--GPU balance using
    one Intel Xeon E5 2620v4 CPU and NVIDIA GeForce GTX 1080 / GTX 1080Ti GPUs.
}
\label{fig:gmx_versions_bench}
\end{figure}

Faster hardware has been a blessing for simulations, but as shown in
\figref{fig:gmx_versions_bench} the improvements in algorithms and software described
here have at least doubled performance {\em for the same hardware} even for older
cost-efficient GPU hardware, and with latest-generation consumer cards the
improvement is almost fourfold over the last five years. 
Given the low-end CPU and high-end GPU combination, new
offload modes bring significant performance improvements when offloading either only PME
or the entire inner iteration to the accelerator.

%%%%%%%%%%%%%%%%%%%%%%%%%%%%%%%%%%%%%%%%%%%%%%%%%
%%%%%%%%%%%%%%%%%%%%%%%%%%%%%%%%%%%%%%%%%%%%%%%%%
%
\begin{figure}[ht]
\begin{center}
\includegraphics[width=0.98\columnwidth]{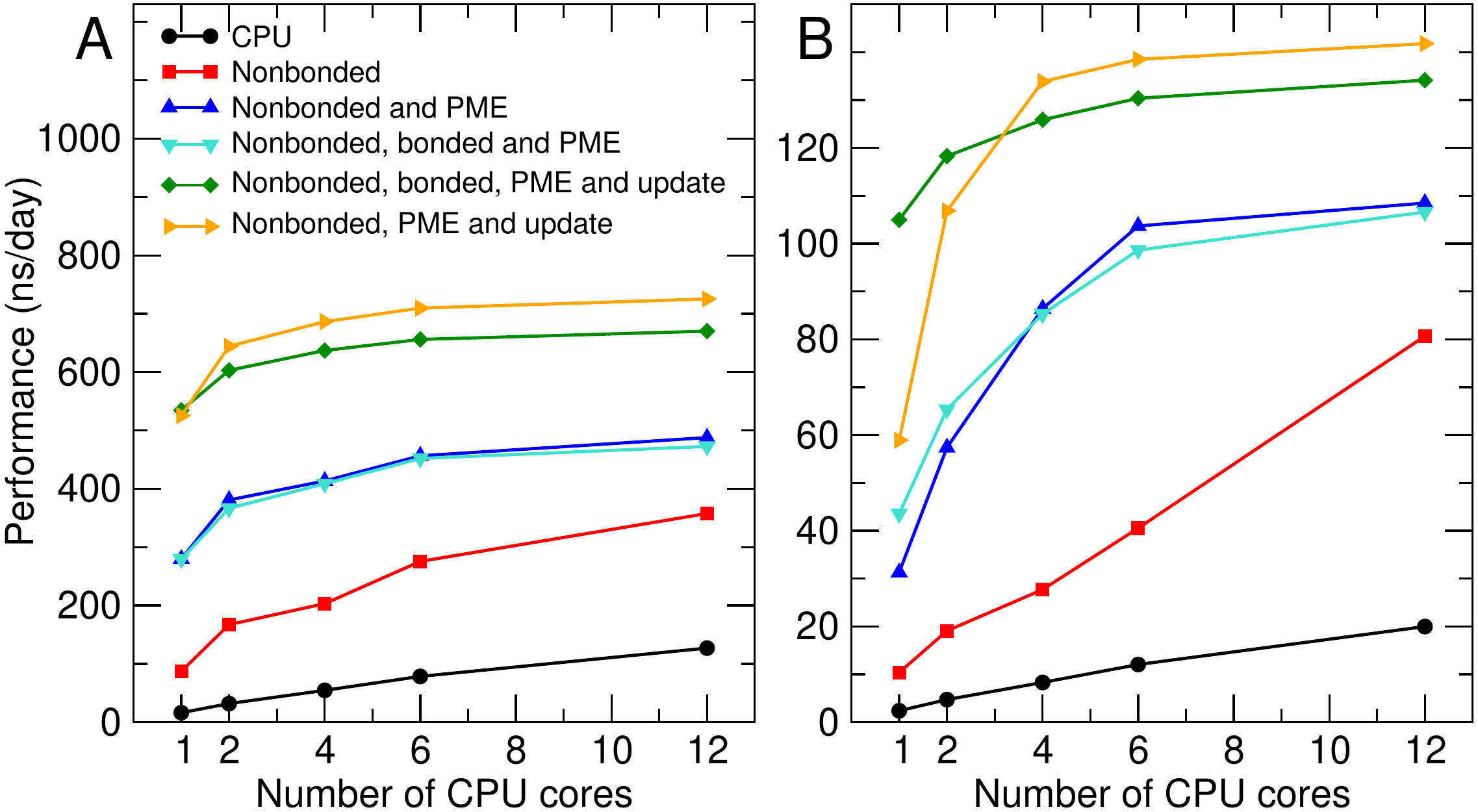}
\end{center}
\caption{Single-GPU performance for (A) RNAse (24k atoms) and (B) GluCl ion channel (142k atoms) systems, both with AMBER99 force-field. Different offload setups are illustrated, with the tasks assigned to the GPU listed in the legend. Hardware: AMD Ryzen 3900X, NVIDIA GeForce RTX 2080 Super.}
\label{fig:single-run-per-gpu}
\end{figure}
Figure~\ref{fig:single-run-per-gpu} shows the impact of different offload
setups for single-GPU runs. As expected, the CPU-only run scales
with the number of cores. When the non-bonded task is offloaded, the performance increases significantly for both systems, but it does not saturate
even when using all cores - indicating
the CPU is oversubscribed. This is
confirmed by the large jump in
performance when the PME task too is offloaded. The GPU now becomes the
bottleneck for computations, and the
curves saturate when enough
CPU cores are used - adding more will not aid performance. 
Consequently, when the bonded forces are also offloaded, there is a performance regression, in particular
for the membrane protein system with lots of torsions. (Figure~\ref{fig:single-run-per-gpu}B). With GPU force tasks taking longer than the CPU force tasks, the data transfers between host and device are no longer effectively overlapping with compute tasks. This is solved by offloading
the entire innermost MD loop,  including coordinate constraining and updating. This leads to another significant jump in performance,  despite the CPU now being mostly idle. To make use of this idle resource, one can move the bonded force evaluation from the GPU back to the CPU. This is beneficial when the entire cycle is
faster than the evaluation of non-bonded and PME forces on the GPU.
For all results the cross-over points will depend on the system, but it is generally faster to evaluate bonded forces on the CPU when apart from very dense systems where only a single CPU core is available per GPU.

\begin{figure}[ht]
\begin{center}
\includegraphics[width=0.98\columnwidth]{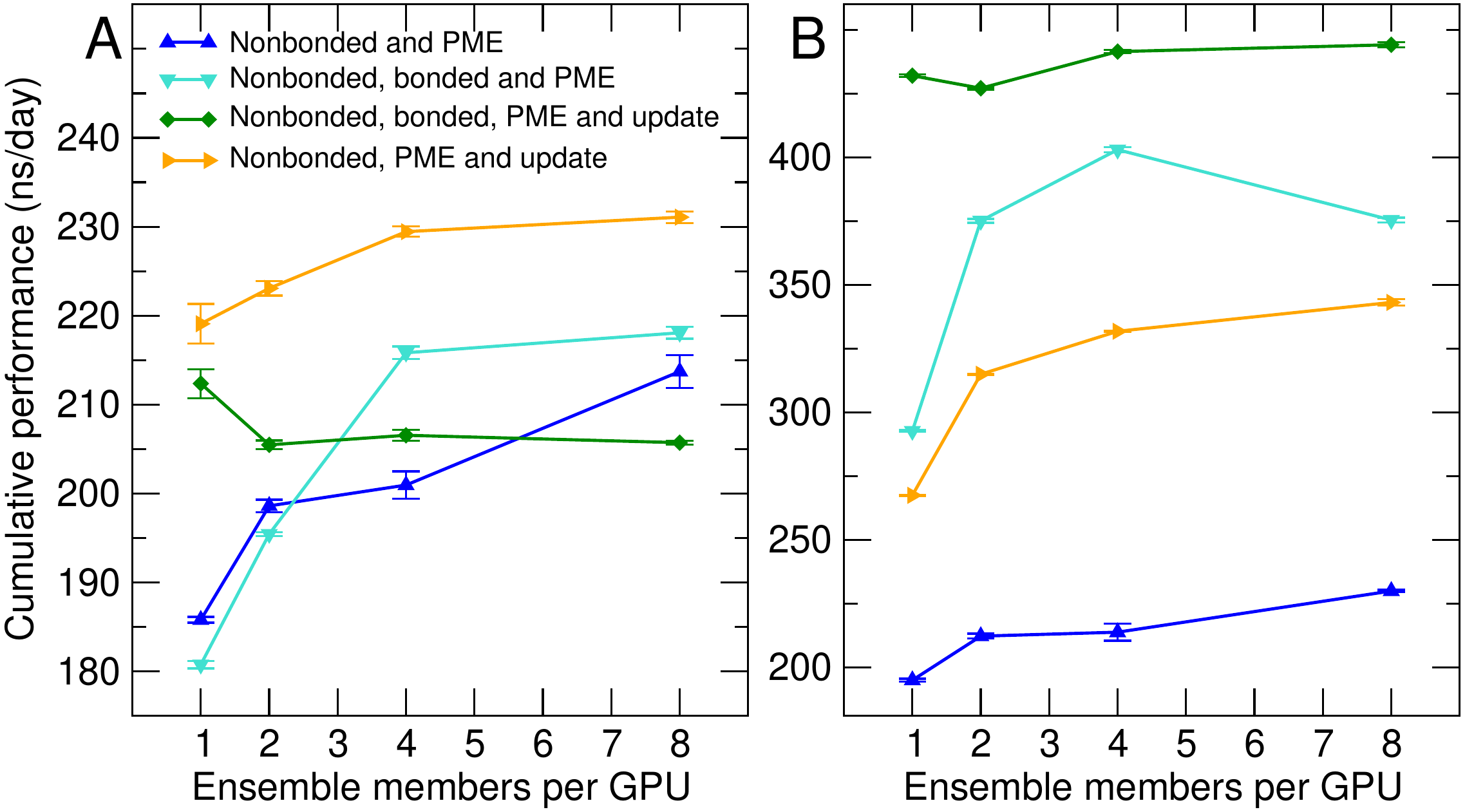}
\end{center}
\caption{Ensemble run cumulative performance as a function of number of Accelerated Weight Histogram walkers executed simultaneously for different offload scenarios. 
The benchmark system is aquaporin (109992 atoms per ensemble member, CHARMM36 force-field). 
The performance was measured on a dual Intel Xeon E5-2620 v4 CPU server with four NVIDIA GeForce GTX 1080 GPUs (A) and four NVIDIA GeForce RTX 2080 GPU (B).}
\label{fig:many-runs-per-gpu}
\end{figure}

Although it is common to run a single simulation per GPU, it is often not the best way of maximizing cumulative throughput since the options to overlap data transfers between CPU and GPU with computational tasks are limited. One way to increase the efficiency even further is to run many uncoupled or loosely coupled trajectories simultaneously on a single GPU. In this case, compute tasks from one trajectory can overlap with the data transfers in another. Figure~\ref{fig:many-runs-per-gpu} shows benchmarks for the case of ensemble simulations using the AWH method \cite{Lindahl2014} on a single node equipped with two CPUs and four GPUs. With medium-performance
GPUs, the best performance is 
achieved when a CPU is used for computing the bonded forces, with everything else evaluated on the GPU
(Figure~\ref{fig:many-runs-per-gpu}A),
and the most efficient throughput
is obtained with four ensemble members
running on each GPU. Using the GPU for the full MD loop is still the second best case when only a single run is executed on the GPU, but for many runs
per GPU there are more options for
overlapping transfer and compute tasks
when using the CPU for updates (integration) and/or
bonded forces. However, for the somewhat older GPUs the difference between the worst- and best-performing cases is only about 25\%. When 
pairing the same older CPUs with 
recent GPUs ((Figure~\ref{fig:many-runs-per-gpu}B), the balance changes appreciably, and
it is no longer justified to use the 
CPUs even for the lightest compute tasks. Performing all tasks on the GPU more than doubles performance compared to leaving updates and bonded forces on the CPU. 
Another advantage is that the throughput does not change significantly with more ensemble members per GPU, which allows for greater flexibility, not to mention the absolute performance will always be highest when only a single ensemble member is assigned to each GPU.

\begin{figure}[ht]
    \center
\includegraphics[width=0.95\columnwidth]{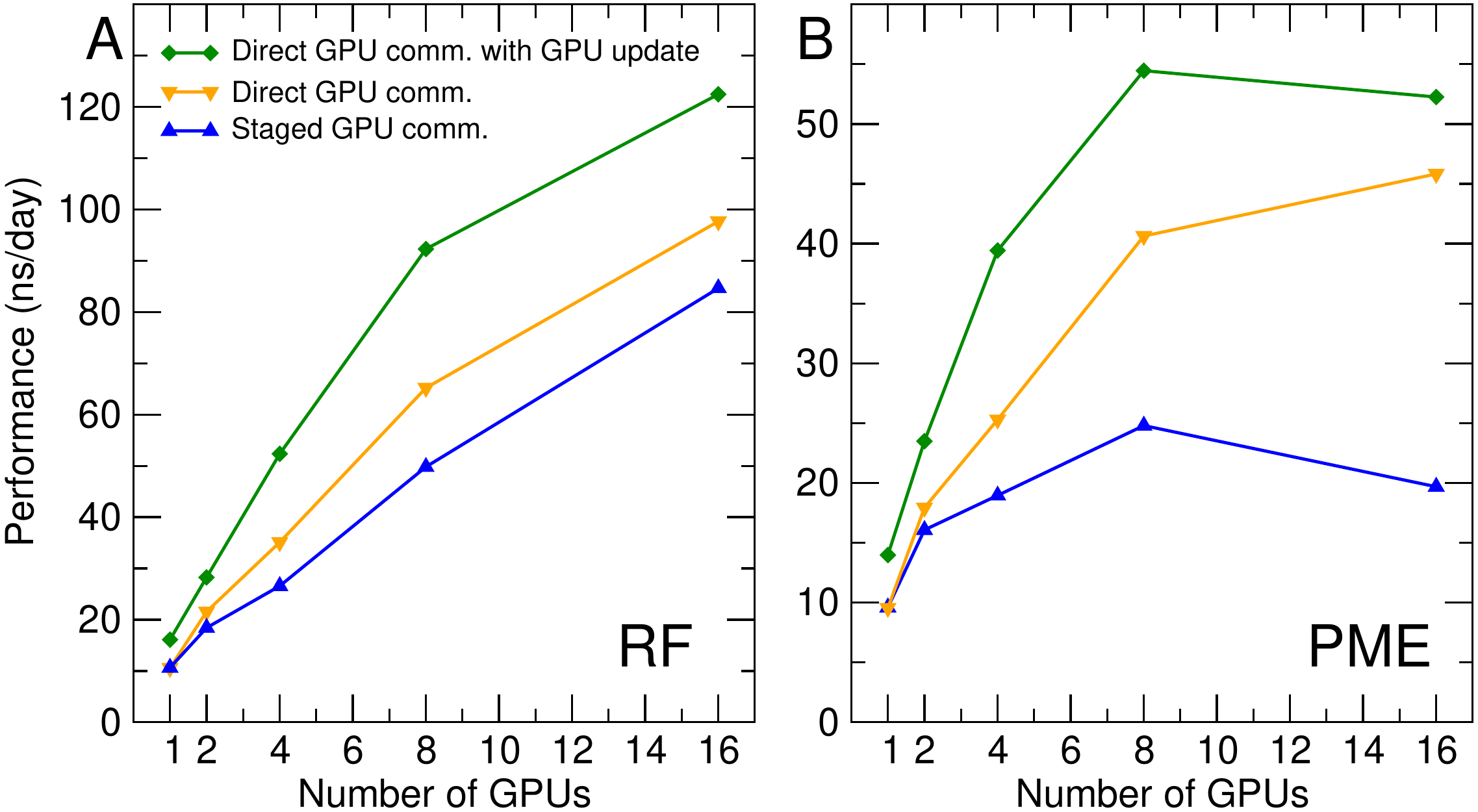}
\caption{
Multi-GPU and multi-node scaling performance STMV benchmark (1M atoms).
Performance when using staged communication through the CPU, direct GPU communication,
and the additional benefit of GPU integration are shown.
Left: When using reaction-field to avoid lattice summation, the 
scaling is excellent and we achieve iteration rates around
\SI{1}{ms} of wallclock time per step over 4 nodes with 4 GPUs each.
Right: With PME, the scaling currently becomes limited by the use a single PME GPUs for offloading, but the absolute performance is high (despite the different scales).
All runs use 1 MPI task per GPU, except the 2-GPU PME runs that use 4 MPI ranks to improve task balance.
}
\label{fig:dgx_scaling}
\end{figure}

Finally, the work on direct GPU communication
now also enables quite efficient multi-GPU scaling combined with outstanding absolute
performance. \figref{fig:dgx_scaling} illustrates the effect of the direct GPU communication optimizations on performance through results from running the 
Satellite tobacco mosaic virus (STMV) benchmark (1M atoms, \SI{2}{fs} steps) on up to four compute nodes, each 
equipped with four NVIDIA Telsa V100 GPUs per node. Intra-node communication uses NVLink and inter-node communication MPI over Infiniband. We believe this configuration is a good match to emerging
next-generation HPC systems, which have a focus on good balance for mixed workloads. When using
reaction-field instead of PME, the scaling is excellent all the way to 16 GPUs. While this is a less common choice for electrostatics, it highlights the efficiency and benefits of the GPU halo exchange combined with GPU update, and shows the performance and scaling possible
when avoiding the challenges with small 3D FFTs, extra communication between direct- and reciprocal-space GPUs, and task imbalance inherent to PME. 
When using PME electrostatics, the relative scaling is good up to 8 GPUs (50\% efficiency) when the
GPU halo exchange is combined with the direct GPU PME task communication,
and there are again clear benefits  from combination with the GPU update path.
Beyond this we are currently limited by the restriction of a single PME GPU when offloading lattice summation,
which becomes a bottleneck both in terms of communication and imbalance in computation. However, we believe the absolute performance of \SI{55}{ns/day} is excellent.

%%%%%%%%%%%%%%%%%%%%%%%%%%%%%%%%%%%%%%%%%%%%%%%%%%%%%%%%%%%%%%%%%%%%%%%%%%%%%%%%%%%%%%%%%%
\section{Discussion}\label{sec:conclusion}
Microsecond-scale simulations have not only become routine but eminently approachable
with commodity hardware. However, it is only the barrier of entry that has been lowered.
Bridging the time-scale gap from hundreds of microseconds to millisecond in single-trajectories still requires special-purpose hardware\cite{Ohmura2014,Grossman2015}.
Nevertheless, general-purpose codes have unique advantages in terms of flexibility, adaptability
and portability to new hardware -- not to mention it is relatively straightforward
to introduce new special algorithms e.g.\ for including experimental constraints in
simulations. There are also great opportunities with using massive-scale resources for
efficient ensemble simulations where the main challenge is cost-efficient generation of
trajectories. Hence, we believe that a combination of new algorithms,
efficient heterogeneous parallelization and large-scale ensemble algorithms will characterize MD in the exascale era - and performance advances in the core MD codes will always multiply the advances obtained from new cleaver sampling algorithms.

The flexibility of the GROMACS engine comes with challenges both for developers
and users. There is a range of options to tune, from algorithmic parameters to
parallelization settings, and many of these are related to fundamental shifts towards much more diverse hardware that the entire MD community has to adapt to.
Our approach has been to provide a broad range of heuristics-based defaults
to ensure good performance out of the box, but by increasingly moving to
tolerance-based settings we aim to both improve quality of simulations and 
make life easier for users.
Still, efficiently using a multitude of different hardware combinations for
either throughput or single long simulations is challenging. As described here, many
of the steps have been automated, but to dynamically decide
e.g.\ the algorithm to resource mapping in a complex compute node (or what code flavor to use)
requires a fully dynamic auto-tuning approach\cite{Glaser2015}.
Developing a robust auto-tuning framework and integrating it with the multi-level load balancers
is especially demanding due to complex feature set and broad use-cases of codes such as GROMACS, but it is something we are working actively on.

To reach performance for which ASICs are needed today,
MD engines need to be capable of $<\SI{100}{\micro \second}$ iterations.
Such iteration rates are possible with GROMACS for small systems, like the 
villin headpice (${\sim} 5000$ atoms)\footnote{$\SI{1.95}{\micro \second/day}$ ($<\SI{88}{\micro \second \per step}$) on an AMD R9-3900X CPU and NVIDIA RTX 2080 SUPER GPU; using \SI{2}{fs} time step and \SI{0.9}{nm} cut-off.}.
Reaching this performance for larger systems like membrane proteins
with hundreds of thousands of atoms will require a range of improvements.
In terms of parallelization in GROMACS, improving the efficiency of GPU task scheduling,
CPU tasking and better overlap of communication are necessary.
When it comes to algorithms, we expect PME to remain the long-range interaction method of choice at low scale,
but the limitations of the 3D FFT many-to-many communication for strong scaling requires a new approach.
With recent extensions to the fast multipole method\cite{Shamshirgar2019}, we expect it to become the algorithm of choice for the largest parallel runs.
Future technological improvements, including faster interconnects and
closer on-chip integration as well as advances in both traditional\cite{Schaffner2018,Yang2019} and coarse-grained reconfigurable architectures\cite{Srivastava2019} could allow getting closer to this performance target.

We expect to see several of these advancements in the future.
In the mean time, we believe the present GROMACS implementation provides a major
step forward in terms of absolute performance as well as relative scaling by being able to use almost
arbitrary-balance combinations of CPU and GPU functional units. It is our hope this will
help enable a wide range of scientific applications on everything from cost-efficient consumer GPU hardware to large HPC resources.

%%%%%%%%%%%%%%%%%%%%%%%%%%%%%%%%%%%%%%%%%%%%%%%%%%%%%%%%%%%%%%%%%%%%%%%%%%%%%%%%%%%%%%%%%%
\section*{Data availability}
The data that support the findings of this study are available in the supplementary material
and in the following repositories:
methodology, topologies, input data and parameters for benchmarks are
published at \url{https://doi.org/10.5281/zenodo.3893789}\cite{SI_GMX_Heterogeneous_JCP_2020};
the source code for multi-GPU scaling is available at \url{https://doi.org/10.5281/zenodo.3890246}\cite{GROMACS_GPUcomm_JCP}.

\begin{acknowledgments}
This work was supported by the Swedish e-Science Research Center, 
the BioExcel CoE (H2020-EINFRA-2015-1-675728), the European Research Council (209825,258980) the Swedish Research Council (2017-04641, 2019-04477), 
and the Swedish Foundation for Strategic Research. NVIDIA, Intel and AMD are kindly acknowledged
for engineering and hardware support. We thank Gaurav Garg (NVIDIA) for CUDA-aware MPI contributions,
Aleksei Iupinov and Roland Schulz (Intel) for heterogeneous parallelization,
Stream HPC for OpenCL contributions, and the project
would not be possible without all contributions from the greater GROMACS community.
\end{acknowledgments}

\bibliography{merged_stripped}

\end{document}